\documentclass{iopart}
\usepackage{iopams}
\usepackage{graphicx}

\begin{document}

\title{Microwave realisation of a periodically driven system}
\author{S.~Gehler$^1$, T.~Tudorovskiy$^{1,2}$, C.~Schindler$^1$, U.~Kuhl$^{1,3}$, H.-J.~St\"{o}ckmann$^1$}
\address{$^1$ Fachbereich Physik der Philipps-Universit\"{a}t Marburg, D-35032 Marburg, Germany}
\address{$^2$ Radboud University Nijmegen, Institute for Molecules and Materials, Heyendaalseweg 135, 6525 AJ Nijmegen, The Netherlands}
\address{$^3$ Laboratoire de Physique de la Mati\`{e}re Condens\'{e}e, CNRS UMR 7336, Universit\'{e} de Nice Sophia-Antipolis, 06108 Nice, France}
\ead{ulrich.kuhl@unice.fr}

\date{\today}

\begin{abstract}
A realisation of a periodically driven microwave system is presented. The principal element of the scheme is a variable capacity, i.e. a varicap, introduced as an element of the resonant circuit. Sideband structures corresponding to different driving signals, have been measured experimentally. In the linear regime we observed sideband structures with specific shapes. The main peculiarities of these shapes can be explained within a semiclassical approximation. A good agreement between experimental data and theoretical expectations has been found.
\end{abstract}

\maketitle

\section{Introduction}
\label{sec:intro}

Periodically driven or kicked quantum systems attract significant interest due to the substantial difference from their classical analogs. Already a simple model of the kicked rotor \cite{cas79a} shows a saturation of the linear raise of energy of a particle as a function of the number of kicks. This phenomenon was called \emph{dynamical localisation} due to its similarity to Anderson localisation \cite{fis82,she87}. There were a few experimental demonstrations using either highly excited hydrogen atoms (Rydberg atom) in strong microwave fields \cite{bay74,gal88,bay89,koc95a,gal94}, ultra-cold atoms within optical traps \cite{moo94,bha99,sad05b,cha08,tal10,lop12} or Bose-Einstein condensates \cite{dan08,tal10}. Due to the new experimental possibilities given by the cold atoms and BEC experiments a renewed theoretical interest in dynamical localization or even more general on periodically driven, e.g., Floquet systems occurred \cite{len11,wan11,tia11,ho12}. However these experiments require a rather complicated experimental setup and usually do not provide a possibility for a large variation of parameters. In contrast to those we propose here a new Floquet setup that is easily accessible and realizable. Its simplest realization already shows quite interesting side-band structures.

Microwave systems have proven to be a very versatile tool to study chaotic and disordered systems (see \cite{stoe99,kuh05b} for reviews). However up to now only time-independent systems have been studied with microwave techniques.

To realise a setup with time-dependent properties using a microwave cavity one has to vary either a physical shape of the cavity (i.e. moving a wall or inset), or its electrical properties. For the realisation of a setup showing dynamical localisation it is necessary to rearrange the whole spectrum. This means that eigenvalues have to be shifted by a distance of the order of the mean level spacing and the frequency of the perturbation should also be of the same order of magnitude. This would mean wall shifts of centimetres with frequencies of several 100\,kHz, which is not realisable mechanically.

Another possibility is a periodic variation of electrical properties of the cavity. One can, for example, induce a dipole antenna attached to a variable capacity (varicap), which can operate up to a few GHz. Such an inset introduces a local perturbation, that can be characterised by a variable scattering length. However experimentally it is difficult to influence the spectrum strong enough using a single local perturbation. Using several varicaps one can try to increase the effect. We have explored the possibility of a local perturbation to realise a setup showing dynamical localisation. Though theoretically such an option may work, experimental results were not satisfactory. We will publish a theory of time-dependent point-like perturbations elsewhere (see also \cite{tud08,tud10}).

If we give up the two-dimensional microwave cavity for a while, we will find a simple possibility to create a microwave setup with a single periodically driven resonance. Indeed, an electric circuit with an integrated varicap obeys our requirements. The capacitance of the varicap depends on the voltage applied to its terminals. Exciting the circuit at some carrier frequency $\omega_c$ and stimulating a varicap with frequency $\Omega$ we receive a response at the frequencies $\omega_c,\,\omega_c\pm \Omega,\,\omega_c\pm 2\Omega,\,\ldots$. Amplitudes of the corresponding harmonics plotted against frequency produce sideband structures.

This simple system already gives nontrivial sideband structures. To understand their main features we introduce a modulation band, i.e. a frequency range between two limiting values corresponding to the largest and the smallest capacities of the varicap. We show that sideband structures can be qualitatively understood by analogy with a semiclassical motion of a particle in a well. The modulation band in this picture corresponds to the classically allowed region. This analogy explains the fast decay of sidebands outside the modulation band and a peculiar behaviour at its border. Basing on the model of the resonant circuit we were able to analyse our experimental results. A good qualitative agreement between the experimental data and the theoretical model provides us an opportunity to report the first successful realisation of the tuned microwave Floquet system. In this paper we will restrict ourself to the linear regime but a nonlinear regime is accessible as well.

Let us now come back to the realisation of a setup showing dynamical localisation. We can think of two possibilities to realise such a system. One can either build an integrated circuit consisting of several coupled devices or to couple our setup to a large cavity providing a large density of states If a single point-like coupling is not sufficient the device or several of them can be coupled to the system by several antennas.

The present paper is structured in the following way: In the next section we will introduce the experimental resonance circuit and characterise its properties by static measurements. In \sref{sec:Motivation} we give hand waving arguments for some of the results. Thereafter we derive theoretically the characteristics of the sideband structures for different types of driving in the single resonance approximation and compare them to numerical calculations. In \sref{sec:ExpResults} we present the experimental Floquet results and compare them to the theoretical predictions (see \sref{sec:LinRespApprox}) and numerics. The paper finishes with a short conclusion.

\section{Experimental setup and its modelling}
\label{sec:ResCircuit}

\begin{figure}
\scalebox{0.8}{
\raisebox{2.0cm}[0pt][0pt]{(a)}
\hspace{-.5cm}
\parbox{3.8cm}{
\includegraphics[width=4.0cm]{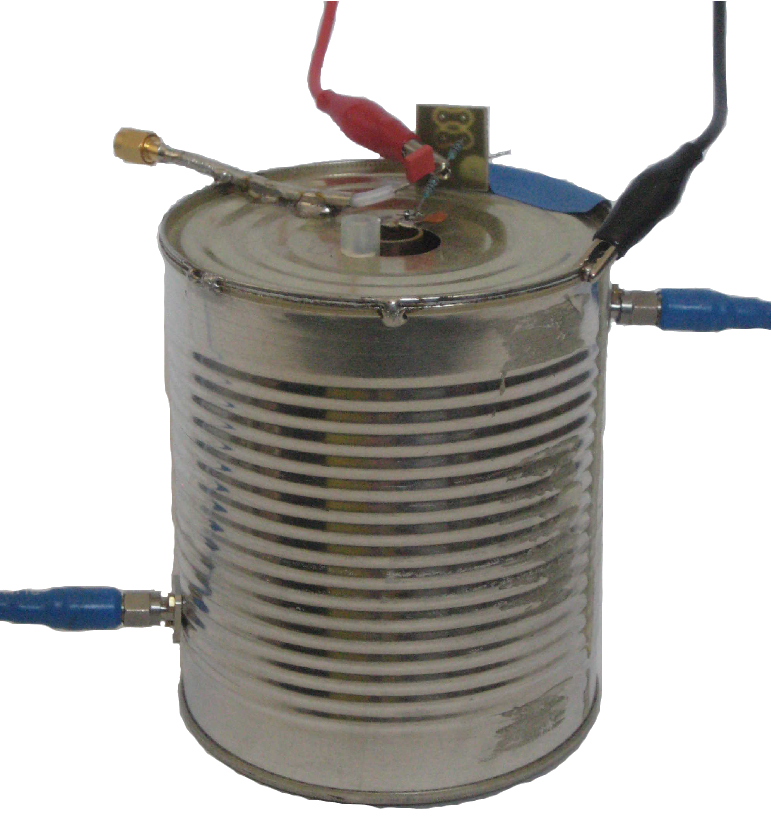}}
\hspace{0.4cm}
\raisebox{2.0cm}[0pt][0pt]{(b)}
\hspace{-1.3cm}
\parbox{6.cm}{
\includegraphics[width=5.5cm]{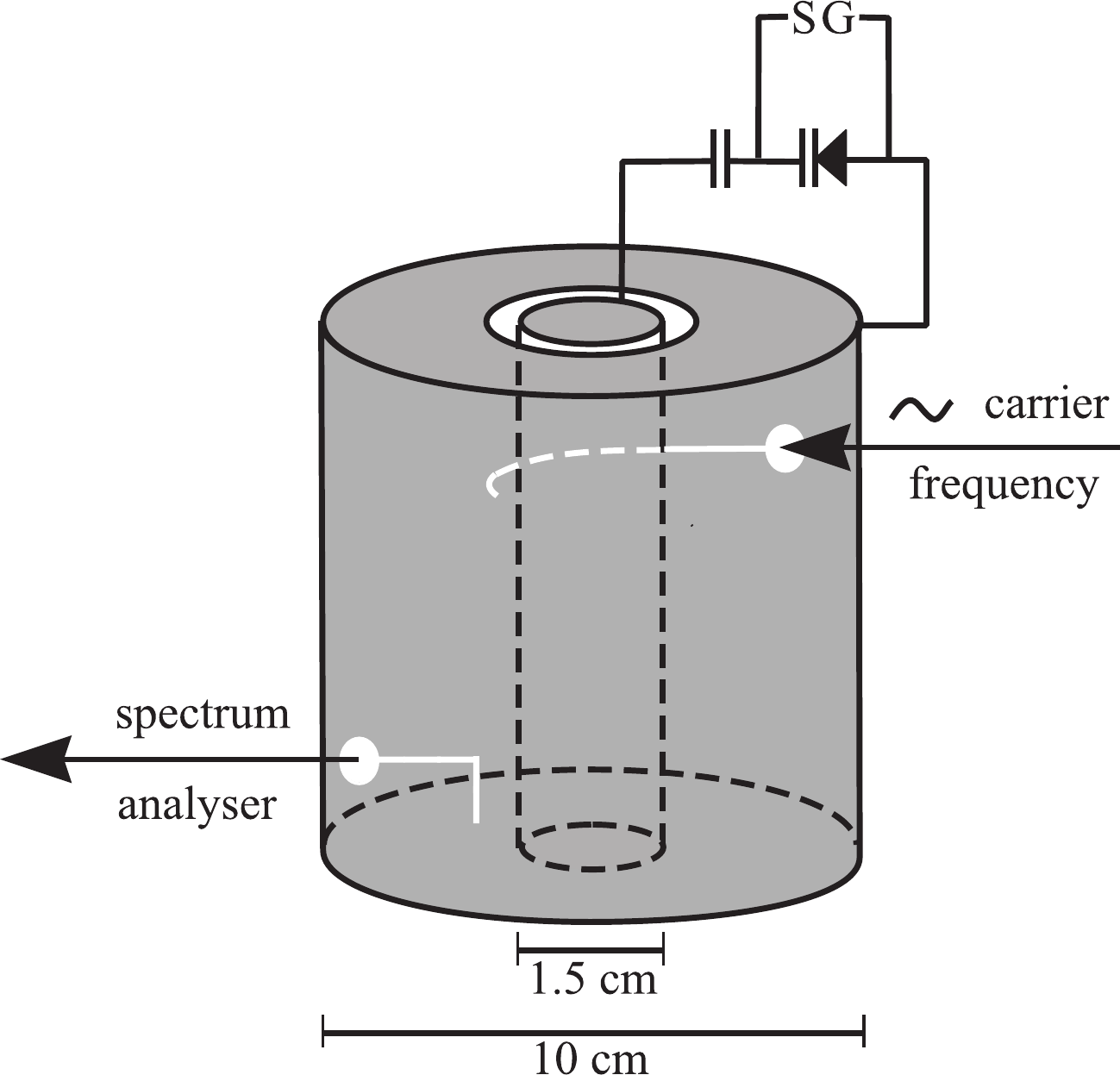}}
\parbox{6.cm}{
\raisebox{1.0cm}[0pt][0pt]{(c)}
\parbox{6.cm}{
\includegraphics[width=5.0cm]{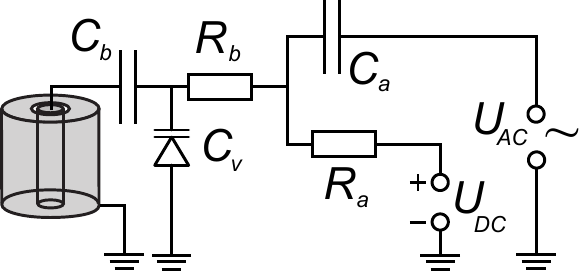}}\\[0.2cm]
\raisebox{1.cm}[0pt][0pt]{(d)}
\parbox{3.8cm}{
\includegraphics[width=5.0cm]{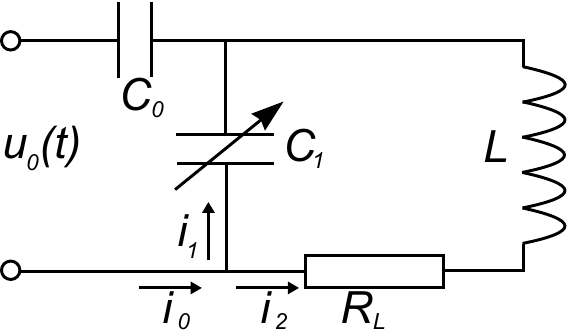}}}}
\caption{\label{fig:Dose}
A photograph of the experimental tin (a), a sketch of the experimental setup (b), a schematic of the varicaps actuation (c), and an approximately equivalent electric circuit of the hole setup (d) are shown. The varicap is shown as the capacitor with an arrowhead. ($C_{\textnormal{v}}$), which reduces in the simplified scheme (d) to a changeable capacitor (capacitor crossed by an arrow ($C_1$)).
}
\end{figure}

The basic requirement on our setup is that it has a resonance of a typical microwave cavity ($\approx 1$\,GHz) which can be shifted electronically via a varicap. In addition one must couple to the system with typical microwave devices. These requirements are ideally met by the tin cup shown in \fref{fig:Dose}(a) in a photograph and \fref{fig:Dose}(b) in a corresponding sketch.

The used tin cup with a metallic core can be described as a cylindrical capacitor with a capacitance of the order of some pF and an inductance of about 45\,nH \cite{zin86,fey77b}. Due to the contact of the core only with the bottom of the tin cup one can describe it by a parallel connection of an inductor and capacitor. Its resonance frequency is of the order of some hundred MHz and a modern UHF-varicap (here Infineon BB833) can change its capacitance in the pF-regime.

To excite and carry out measurements on our system we placed two antennas (white in \fref{fig:Dose}(b)) at the tin cup. The upper antenna is forming a capacitor with the core of the tin cup, thus coupling capacitively. This is in contrast to the lower antenna, which is coupling inductively to the magnetic field of the core, due to the parallelism of its last part and the core. Because of these two different couplings there is no direct communication between the two antennas.

\Fref{fig:Dose}(c) shows the circuit in more detail, including the capacities $C_a$ = 100\,nF, $C_b$ = 22\,pF, and the resistances $R_a$ = $R_b$ = 47\,k$\Omega$.

A simplified equivalent circuit diagram for the whole setup is shown in \fref{fig:Dose}(d). Due to the description of the tin cup as a parallel circuit $L$, $C_0$, $C_1$, and $R_L$ depend slightly on frequency. In the following we neglect these dependencies. Feeding a signal $u_0(t)$ with an angular carrier frequency $\omega_c= 2 \pi \nu_c$ from the source (Anritsu 68047C) to the upper antenna, we feed an external sinusoidal AC voltage to the circuit. Another voltage, either a DC voltage ($U_{DC}$ - for the static measurement) or an AC voltage ($U_{AC}$) with frequency $\omega_d= 2 \pi \nu_d$, is applied to the terminals of the varicap. For the major measurements we used a signal generator (HP 33120A, constant bias voltage adjustable) to achieve a periodic modulation of the resonance frequency. Via the lower antenna the current through the inductor is measured by the spectrum analyser (Rohde\&Schwarz FSU), which measures the intensities of the different frequency components of the applied signal.

Let us describe the electric circuit shown in the \fref{fig:Dose}(d). By $i_0$, $i_1$, $i_2$ we denote currents on the corresponding transmission lines. The application of the Kirchhoff's first law gives
\begin{equation}
 i_0=i_1+i_2.
 \label{eq::1}
\end{equation}
The charging current at the capacitor $C_0$ is $i_0=\dot q_0$, where $q_0$ is the charge of the capacitor, and $i_1=\dot q_1$, where $q_1$ is the charge of the capacitor $C_1(t)$.

The application of the Kirchhoff's second law for the two loops gives
\begin{eqnarray}
 L \frac{d i_2}{dt}+R_L i_2+q_0/C_0=u_0(t),\label{eq::2}\\
 q_1/C_1(t)+q_0/C_0=u_0(t).
 \label{eq::3}
\end{eqnarray}
Integrating \eref{eq::1} we obtain
\begin{equation}
 q_1=q_0-q_2,\quad q_2=\int i_2(t)\,dt.
 \label{eq::4}
\end{equation}
On the other hand, we have from \eref{eq::3}:
\begin{equation}
 q_1=u_0(t)C_1(t)-q_0 C_1(t)/C_0.
 \label{eq::5}
\end{equation}
Equating \eref{eq::4} and \eref{eq::5} we find
\begin{equation}
 q_0=\frac{C_0 C_1(t)}{C_0+C_1(t)}u_0(t)+\frac{C_0}{C_0+C_1(t)}q_2.
 \label{eq::7}
\end{equation}
Substituting \eref{eq::7} into \eref{eq::2} we obtain
\begin{equation}
 L \ddot q_2+R_L \dot q_2+\frac{1}{C_0+C_1(t)}q_2=\frac{C_0}{C_0+C_1(t)}u_0(t).\label{eq::9}
\end{equation}

Experimentally the change of the varicap capacity is relatively small, i.e. $C_1(t)=C_1+\delta C_1(t)$, $|\delta C_1(t)|\ll C_1$. Introducing the notation $C=C_0+C_1$ we rewrite \eref{eq::9} as
\begin{equation}
 L \ddot q_2+R_L \dot q_2+\frac{1}{C+\delta C_1(t)}q_2=
 \frac{C_0}{C+\delta C_1(t)}u_0(t).\label{eq::9-1}
\end{equation}

We consider $u_0(t)=u_0 e^{-i\omega_c t}$, where $\omega_c$ is the carrier frequency. We restrict ourselves to the case when the sideband structure is generated by a single driven resonance. For a constant capacity \eref{eq::9-1} is nothing but the equation for the driven harmonic oscillator with its long-time solution given by $q_2(t)=\mathrm{const.}\times u_0(t)$, but for a time-dependent capacity the amplitude of the oscillation becomes time-dependent, too. This suggests the ansatz $q_2(t)=-u_0e^{-i\omega_c t}C_0 f(t)/2$, where $f(t)$ describes the amplitude modulation. $f(t)$ obeys the equation
\begin{equation}
\fl
\left(\frac{\omega_c^2}{\omega_0^2}-\frac{C}{C+\delta C_1(t)}+\frac{2i\gamma\omega_c}{\omega_0^2}\right)f+
 \frac{2(i\omega_c-\gamma)}{\omega_0^2}\dot f-\frac{\ddot f}{\omega_0^2}=\frac{2C}{C+\delta C_1(t)}.
 \label{eq::10}
\end{equation}
where $\gamma=R_L/2L$ and $\omega_0^2=1/LC$. The first two terms in the brackets can be written as
\begin{equation}
\frac{\omega_c^2}{\omega_0^2}-\frac{C}{C+\delta C_1(t)}=\frac{\omega_c^2-\omega_0^2}{\omega_0^2}+\frac{\delta C_1(t)}{C+\delta C_1(t)}
\end{equation}
Neglecting $\delta C_1(t)$ as compared to $C$ we find
\begin{equation}
 \left(\frac{2(\omega_c-\omega_0)}{\omega_0}+
 \frac{\delta C_1(t)}{C}+\frac{2i\gamma}{\omega_0}\right)f+
 \frac{2i}{\omega_0}\dot f-\frac{\ddot f}{\omega_0^2}=2.
 \label{eq::11}
\end{equation}
Let us estimate the magnitude of $f$. Neglecting derivatives we find $f\sim\omega_0/\Delta\omega$, where $\Delta\omega=\omega_0(\delta C_1/C_1)$ is the width of the modulation band. From the requirement that the term, containing the first derivative, is comparable to $\Delta\omega f/\omega_0$ we conclude that $\dot f\sim\Delta\omega f$. Hence $\ddot f\sim (\Delta\omega)^2 f\sim (\Delta\omega)\omega_0$ and $\ddot f/\omega_0^2\sim (\Delta\omega)/\omega_0$. Therefore the term containing the second derivative in \eref{eq::11} can be neglected. Finally we obtain
\begin{equation}
 \left(i\frac{d}{dt}+\omega_c-\omega_0+d_1(t)+i\gamma\right)f
 =\omega_0, \quad d_1(t)=\frac{\delta C_1(t)}{2C}\omega_0.
 \label{eq::12}
\end{equation}

\begin{figure}
\begin{center}
\raisebox{1.5cm}[0pt][0pt]{(a)}\hspace*{-0.35cm}
\parbox{6cm}{\includegraphics[width=6cm]{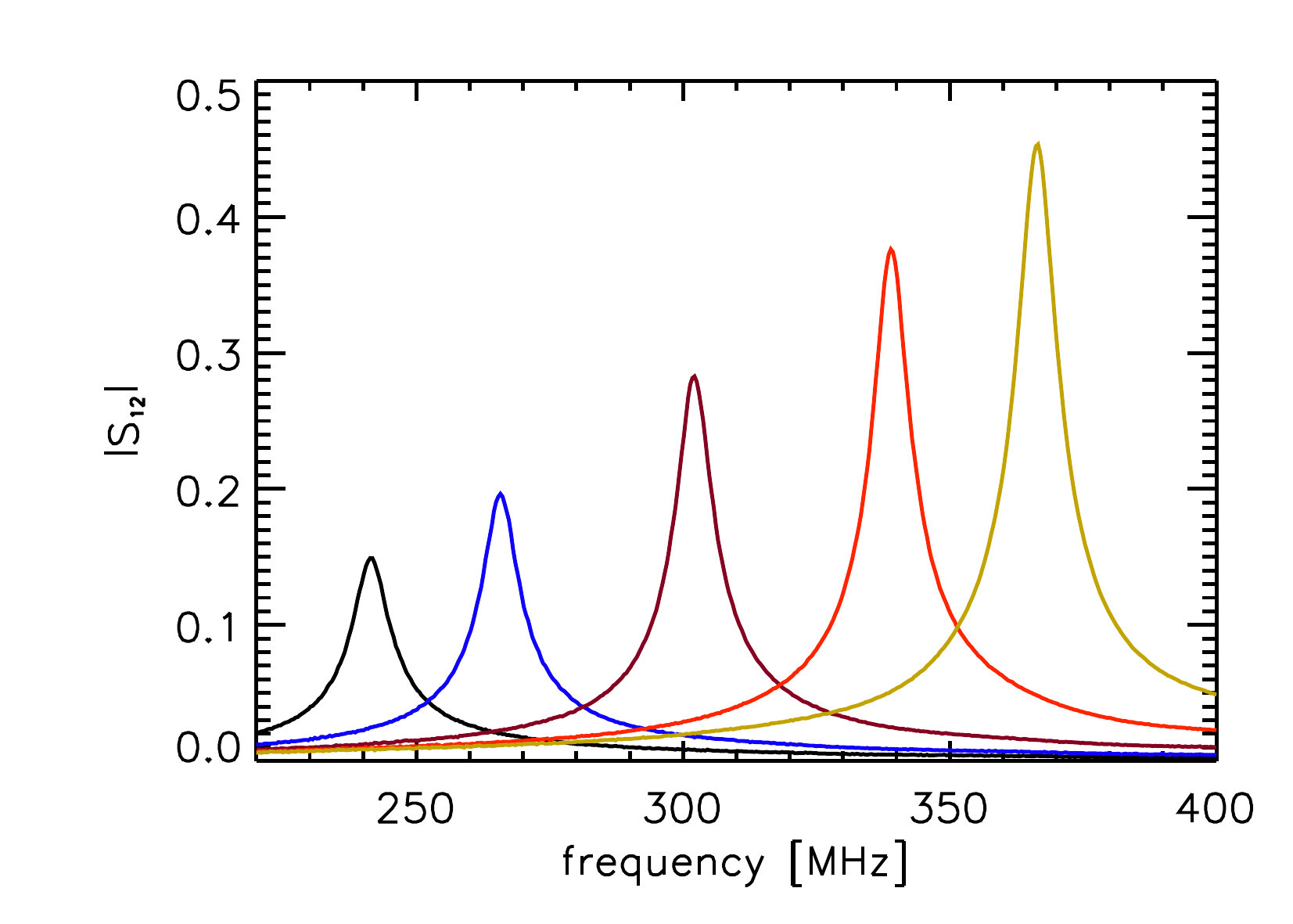}}
\end{center}
\begin{center}
\mbox{
\raisebox{1.5cm}[0pt][0pt]{(b)}\hspace*{-0.35cm}
\parbox{6cm}{\includegraphics[width=6cm]{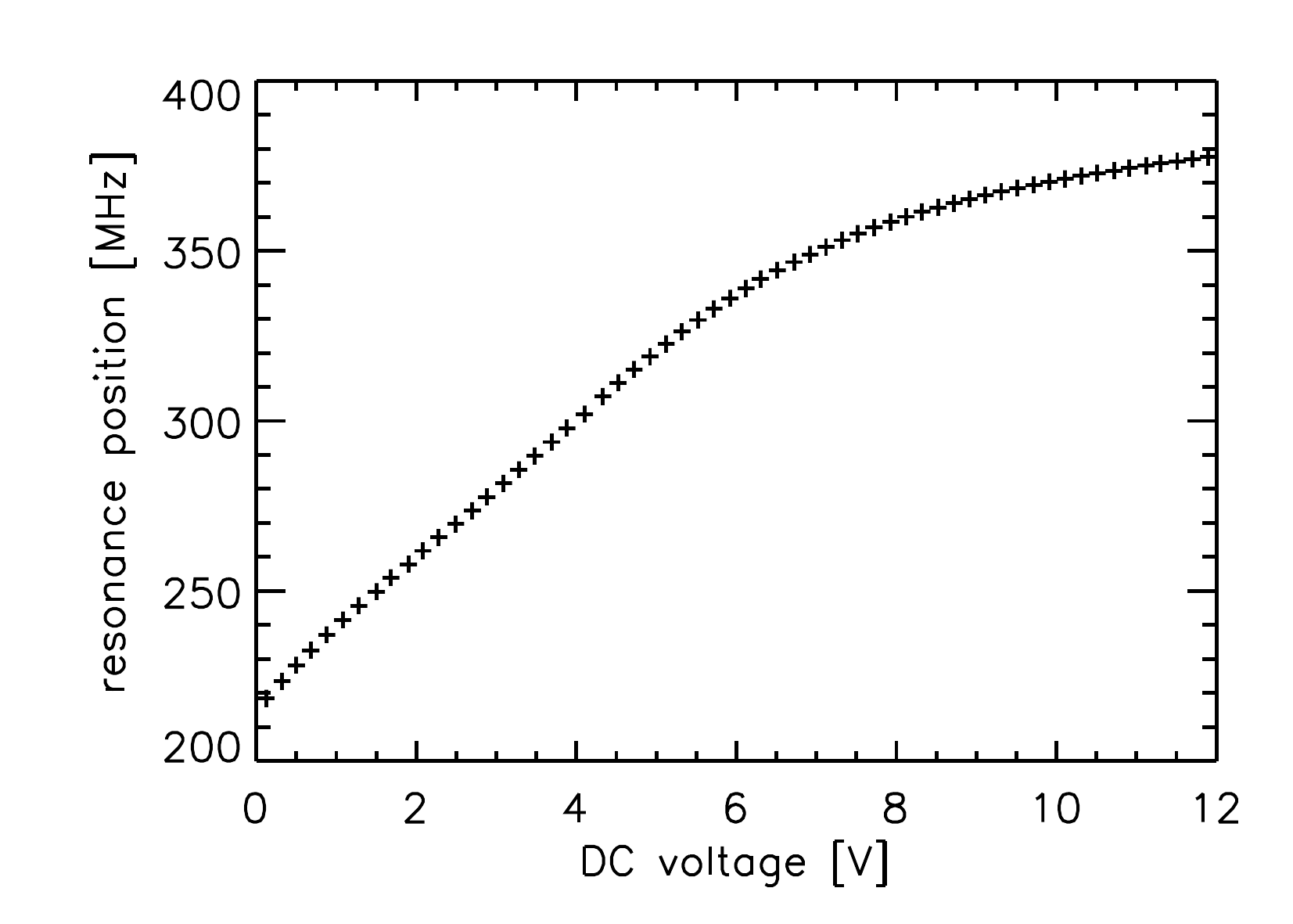}}
\raisebox{1.5cm}[0pt][0pt]{(c)}\hspace*{-0.35cm}
\parbox{6cm}{\includegraphics[width=6cm]{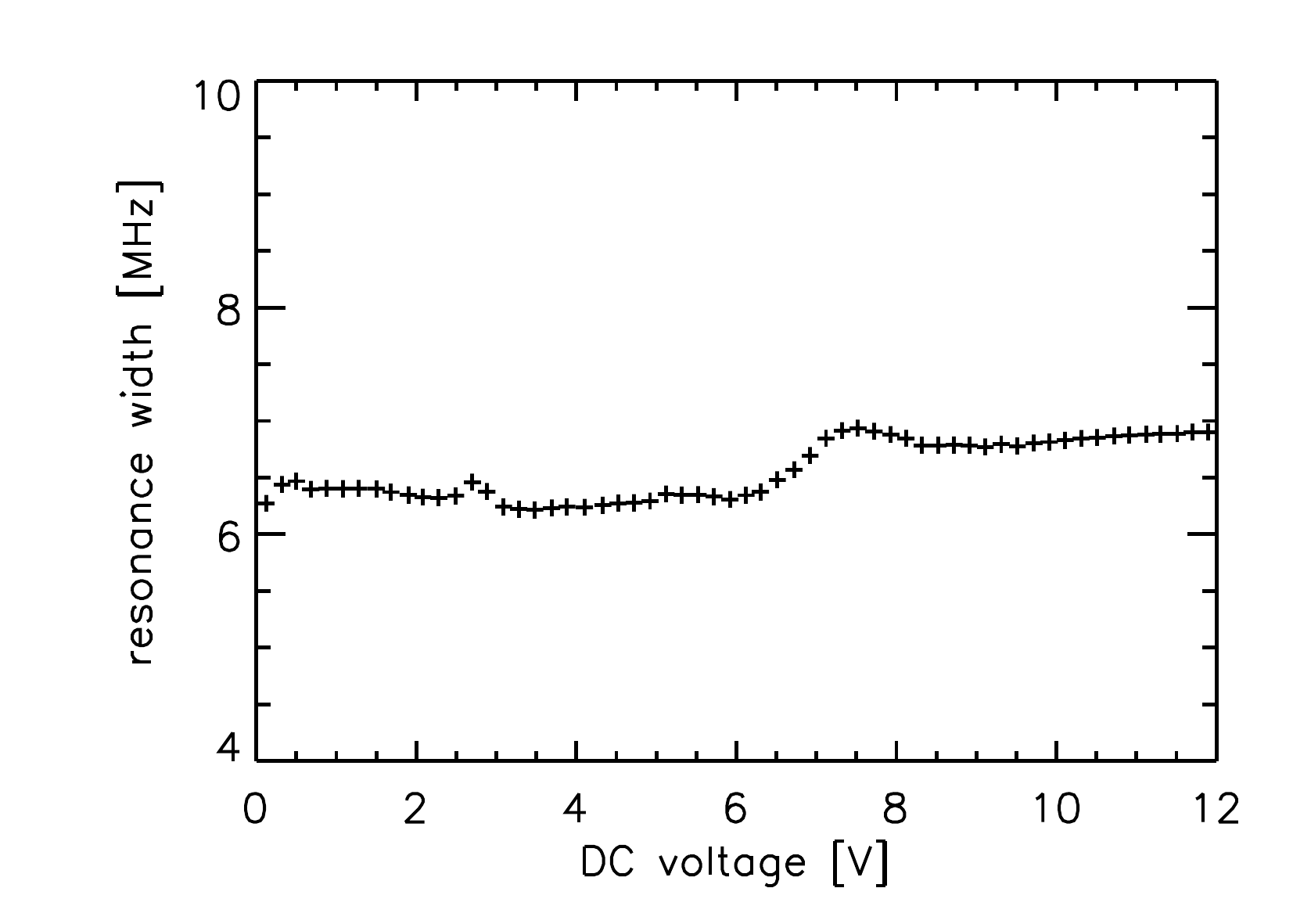}}}
\end{center}
\caption{\label{fig:res-pos}
(a) Experimentally measured resonances for different DC voltages $U_{DC}$ applied to the diode. Voltage values from the left to the right: $U_{DC}$= 1\,V, 2.2\,V, 4\,V, 6\,V, and 9\,V. Resonance position and width as a function of the applied DC voltage are shown in (b) and (c).}
\end{figure}

Experimentally we measure the current $i_2=\dot q_2$ flowing through the coil. In the considered case we can use the approximation
\begin{equation}
i_2=\frac{i}{2}\omega_c u_0 C_0 e^{-i\omega_c t}f(t).
\end{equation}
Thus the spectrum measured by the spectrum analyser is given by the Fourier transformation of $f(t)$.

To characterise the values of the devices in \fref{fig:Dose}(d),  measurements with a time-independent $d_1(t)=d_1$ were performed. In this static measurements the amplitude $f$ as a function of the carrier frequency $\omega_c$ has a resonant behaviour
\begin{equation}\label{eq:f}
f=\frac{\omega_0}{\omega_c-\omega_0+d_1+i\gamma}.
\end{equation}
The power of the transmitted signal is proportional to $|f|^2$, i.e. it is described by a Lorentzian function. The dependencies of the transmitted power on the carrier frequency for different values $U$ of the voltage on the terminals of the varicap are shown in \fref{fig:res-pos}(a). The increase of the resonance height is due to the $\omega_0$ dependence in the nominator in \eref{eq:f}. The resonance frequency is given by $\omega_0-d_1$. In \fref{fig:res-pos}(b) the dependence of the resonance frequency is shown, from which the dependence of $d_1=d_{DC}(U)$ can be extracted. The width of the resonance, associated to $2\gamma=2\gamma_{DC}(U)$, depends only weakly on the applied voltage as can be seen from \fref{fig:res-pos}(c).

Replacing $U$ by $U(t)$ in the case of the dynamic measurement we obtain
\begin{equation}
d_1(t)=d_{DC}\bigl(U(t)\bigr), \qquad \gamma(t)=\gamma_{DC}\bigl(U(t)\bigr).
\end{equation}

\section{Preliminaries}
\label{sec:Motivation}

Let us take a heuristic look on the introduced setup. If one applies an AC-voltage with driving frequency $\omega_d$ at the terminals of the varicap, then the properties of the circuit are changing in time periodically. The shape of the AC-signal can be chosen arbitrarily, the shape restrictions are imposed by the available signal generator.

\begin{figure}
\raisebox{1.5cm}[0pt][0pt]{(a)}
\parbox{6cm}{\includegraphics[width=6cm]{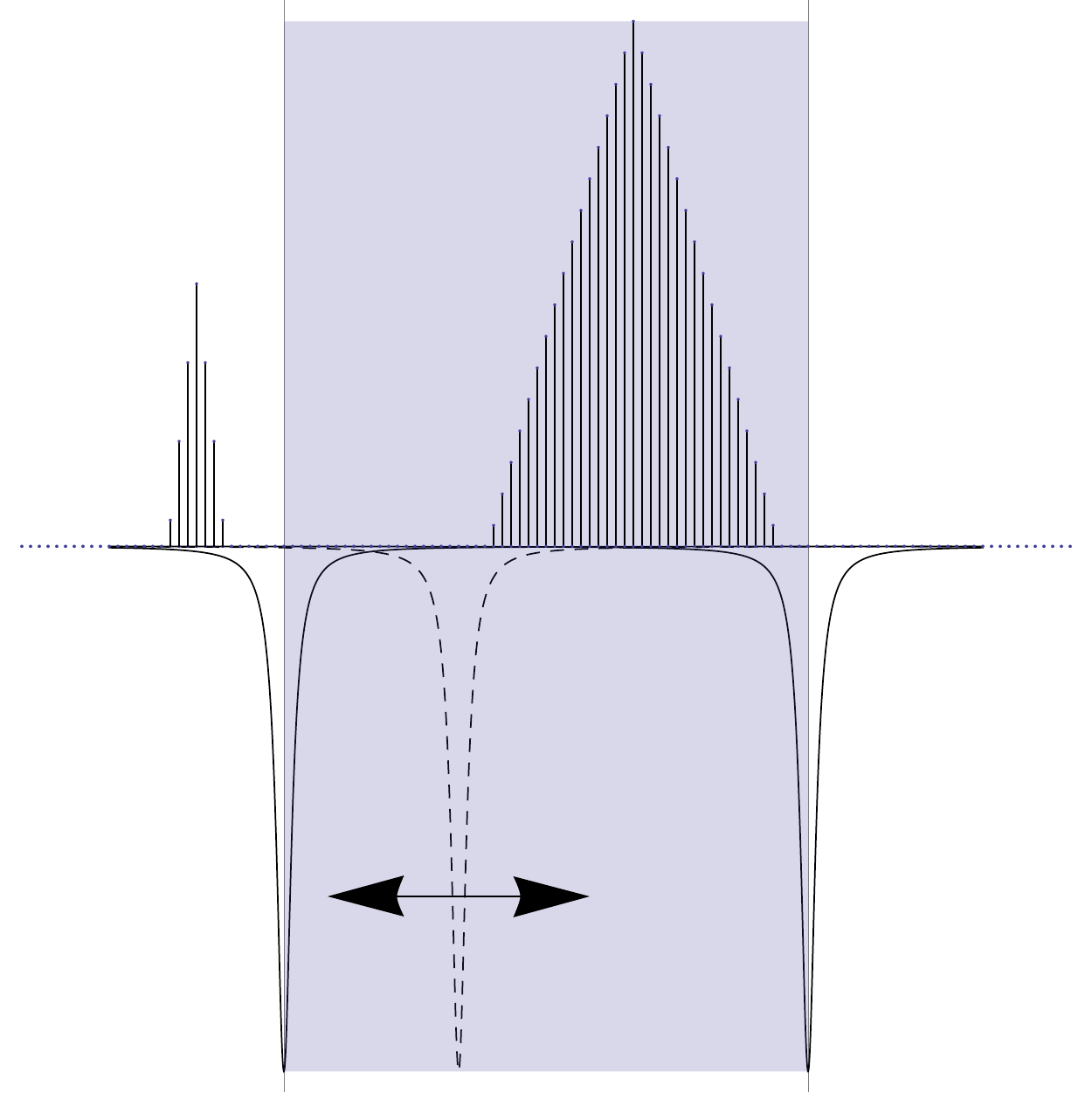}}
\raisebox{1.5cm}[0pt][0pt]{(b)}
\parbox{6cm}{\includegraphics[width=6cm]{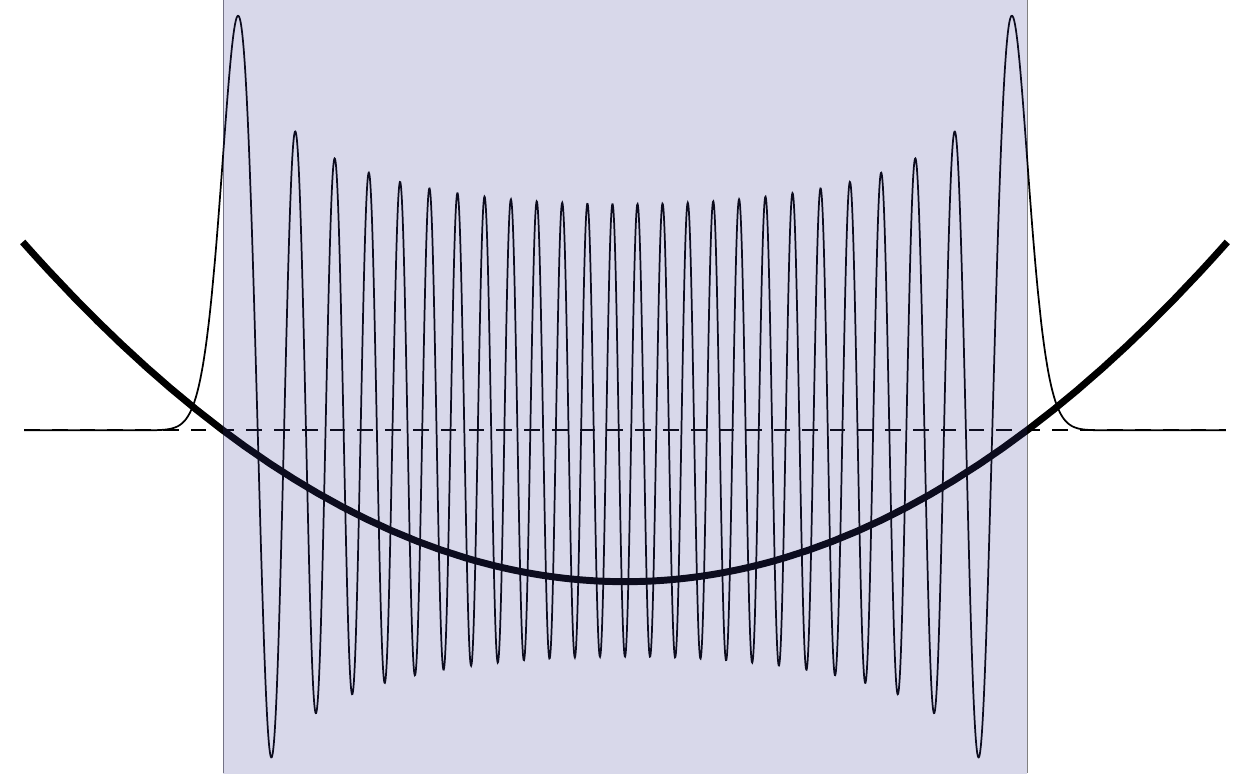}}
\hspace{.5cm}
\caption{\label{fig:intro}
(a) Schematic illustration of the expected sideband structure. The upper part corresponds to the sideband structure of for two different AC driving frequencies, one outside and the other one within the allowed region. Below the DC resonance structure for the maximal and minimal AC voltage is shown, defining the classically allowed range (shaded in blue). (b) A typical eigenfunction of the harmonic oscillator.}
\end{figure}

The periodically driven circuit is excited at a certain carrier frequency $\omega_c$. This means an additional AC-voltage applied to the circuit. $\omega_c$ is assumed to be close to the resonance frequency of the circuit. Since the system is changing in time periodically, it will respond at the frequencies $\omega_c,\,\omega_c\pm \omega_d,\,\omega_c\pm 2\omega_d,\,\ldots$ leading to a sideband structure. The amplitudes of sidebands can be detected by a spectrum analyser (see \fref{fig:intro}). The analysis of these structures is the main subject of the paper.

The circuit shown in \fref{fig:Dose}(c) assumes a linear dependence of the response on the strength of the carrier signal. However we found that a seemingly weak signal can already lead to a nonlinear dependence of the shape of sideband structures on the applied power. To avoid complications associated with non-linear characteristics of the setup we have lowered the strength of the signal to $-20$\,dBm. If one now slightly changes the power ($\sim 1$\,dBm), the amplitudes of all peaks will be altered linearly giving the same sideband structure structure.

What do we expect in the linear regime? In \fref{fig:intro}(a) we have summarised schematically what happens when we change the voltage on the varicap terminals in the static regime between two limiting values (lower half-plane) and in the dynamic regime (upper half-plane). In the static measurement of the transmitted signal we see the peak at the resonance frequency which depends on the voltage on the terminals. Changing the voltage we shift the resonance. The modulation band between two limiting positions  of the resonance peak (filled range in \fref{fig:intro}(a)), corresponding to two limiting values of the voltage on the terminals, forms the ``classically allowed'' range of frequencies. Analogously the range outside this interval is called ``classically forbidden''. These notations are motivated by the resemblance to the wavefunctions of a harmonic oscillator potential in the semiclassical regime (see \fref{fig:intro}(b)). This analogy will be explained below in \sref{ssec:semiclassics}.

Let us turn to the sideband structures observed in the dynamic regime. Periodically changing the voltage within the range corresponding to the ``classically allowed'' frequencies, and exciting the setup at some carrier frequency, we generate a transmitted signal. The characteristic Fourier spectrum of the transmitted signal is shown in \fref{fig:intro}(a), upper part. As long as the carrier frequency is taken within the classically allowed region, a large number of sidebands is observed. The situation is qualitatively different if the carrier frequency is taken in the classically forbidden region. Now there is only a small number of sidebands decaying rapidly with the sideband number. For the carrier frequency taken close to one of the turning points, peculiar asymmetric sideband structures are observed. All these features will be discussed in detail in \sref{sec:ExpResults}.

Let us go on with a more detailed theoretical description.

\section{Sidebands generated by a single driven resonance}
\label{sec:LinRespApprox}

In this section we consider sideband structures generated by a single driven resonance within approximation \eref{eq::12}. The equation is inhomogeneous, i.e. it has a non-zero right hand side associated with a source. We introduce a solution
\begin{equation}
\psi(t)=\exp\left(i(\omega_c-\omega_0)t-\gamma t+i\int_0^t dt'd_1(t')\right)
\end{equation}
of the homogeneous equation
\begin{equation}
 \left(i\frac{d}{dt}+\omega_c-\omega_0+d_1(t)+i\gamma\right)\psi=0.
 \label{eq::f0}
\end{equation}
Then the periodic solution of \eref{eq::12} reads
\begin{eqnarray}
f(t)=-i\omega_0 \psi(t)\int_{-\infty}^{t}dt' \psi^{-1}(t').
\end{eqnarray}
Now we can expand $f(t)$ in the Fourier series. To this end we write
\begin{eqnarray}
\exp\left(i\int_0^t dt'd_1(t')\right)=\sum_{n=-\infty}^\infty a_n \exp(-in\Omega t),\nonumber\\
a_n=\frac{\Omega}{2\pi}\int_{-T/2}^{T/2} dt \exp\left(in\Omega t+i\int_0^t dt'd_1(t')\right),
\label{eq:a_n}
\end{eqnarray}
where $\Omega=2\pi/T$. Then
\begin{eqnarray}
f(t)=\omega_0\sum_{n,m=-\infty}^\infty a_n a_m^*\frac{e^{-i(n-m)\Omega t}}{\omega_c-\omega_0-m\Omega+i\gamma}=\sum_{n=-\infty}^\infty f_n e^{-i n\Omega t},\\
f_n=\omega_0\sum_{m=-\infty}^\infty
\frac{a_{n+m} a_{m}^*}{\omega_c-\omega_0-m\Omega+i\gamma}.
\label{eq:f_n}
\end{eqnarray}
The coefficients $f_n$ correspond to the amplitudes of sidebands shown in \fref{fig:intro}(a). In the experiment the Fourier analysis was performed by a spectrum analyser, as was mentioned above, yielding, however, only the moduli of the sideband amplitudes but not their phases.

\subsection{Calculation of harmonics}
\label{ssec:HarmonicCalc}

To determine $f_n$ from equation \eref{eq:f_n} one has first to compute coefficients $a_n$ from \eref{eq:a_n}. Let us consider two examples, where $a_n$ can be calculated exactly.

For the sinusoidal driving
\begin{equation}
d_1(t)=Z\cos(\Omega t)
\label{eq:sinus}
\end{equation}
we obtain
\begin{equation}\label{anSinuasoidal}
a_n=(-1)^n J_n\Bigl(Z/\Omega\Bigr),
\end{equation}
where $J_n$ is the Bessel function of the first kind. In case of rectangular driving
\begin{equation}
d_1(t)=\left\{
\begin{array}{ll}
  -Z, & \textrm{for } -T/2<t<0,\\
   Z, & \textrm{for } 0<t<T/2.
\end{array}
\right.
\end{equation}
we find
\begin{equation}\label{anRectangular}
a_n=\left\{
\begin{array}{lll}
\displaystyle
\frac{Z}{i\pi\Omega}\frac{1-(-1)^n e^{i\pi Z/\Omega}}{n^2-(Z/\Omega)^2}, & \textrm{if} &n\neq \pm Z/\Omega.\\
\displaystyle
1/2,  &\textrm{if} & n = \pm Z/\Omega.
\end{array}\right.
\end{equation}

\begin{figure}
\mbox{
\parbox{\textwidth}{
\includegraphics[width=.45\textwidth]{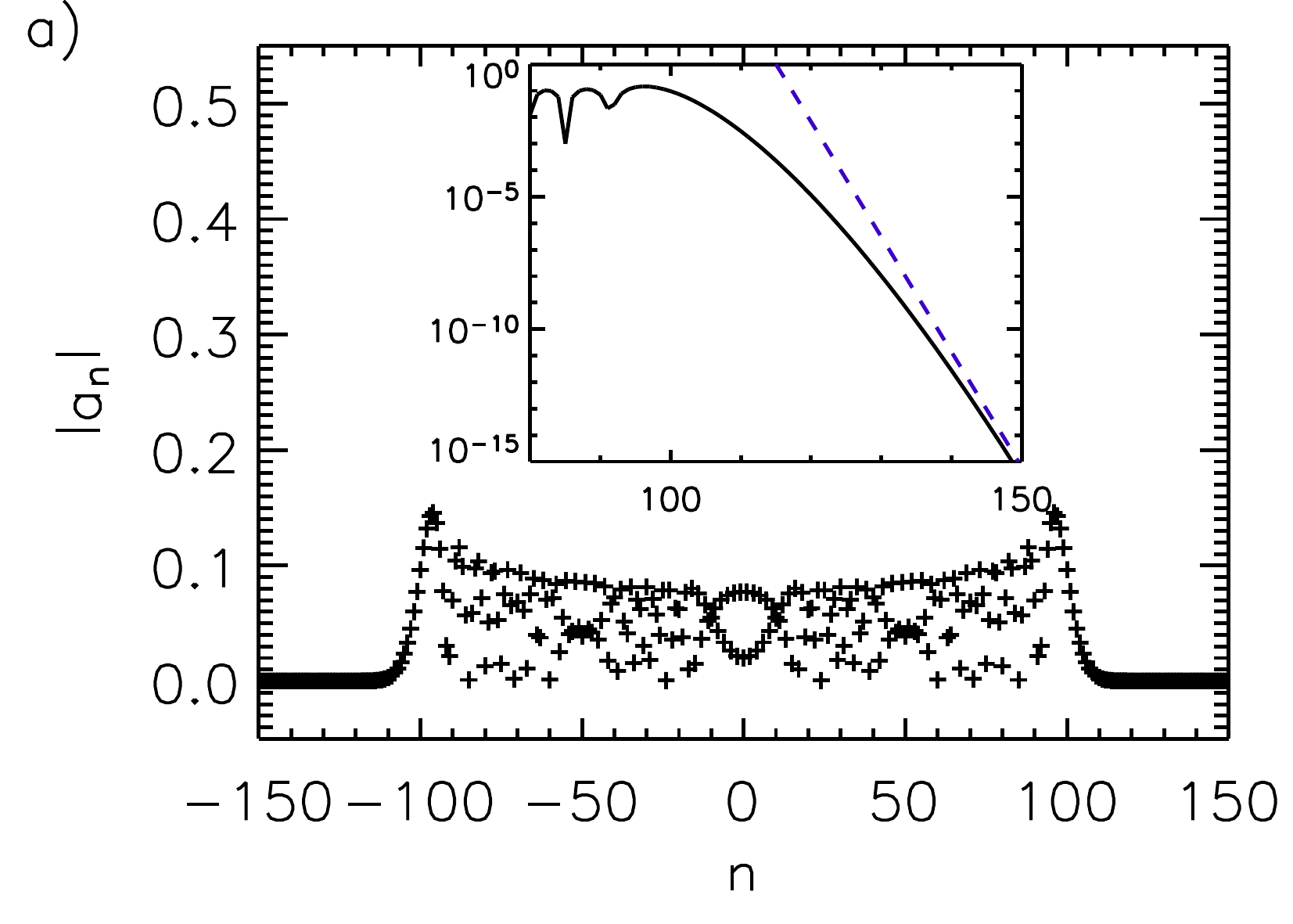}\qquad
\includegraphics[width=.45\textwidth]{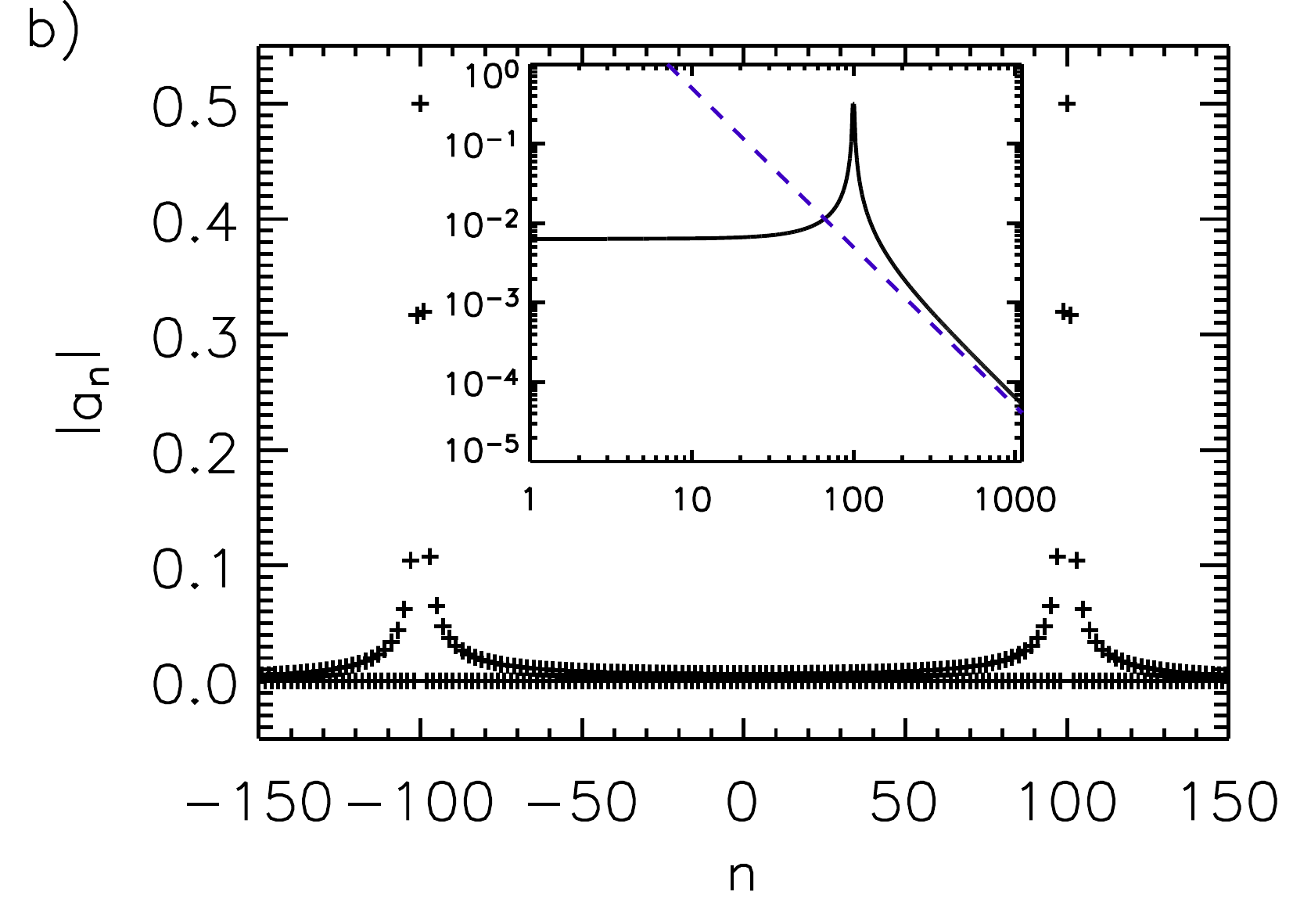}\\
\includegraphics[width=.45\textwidth]{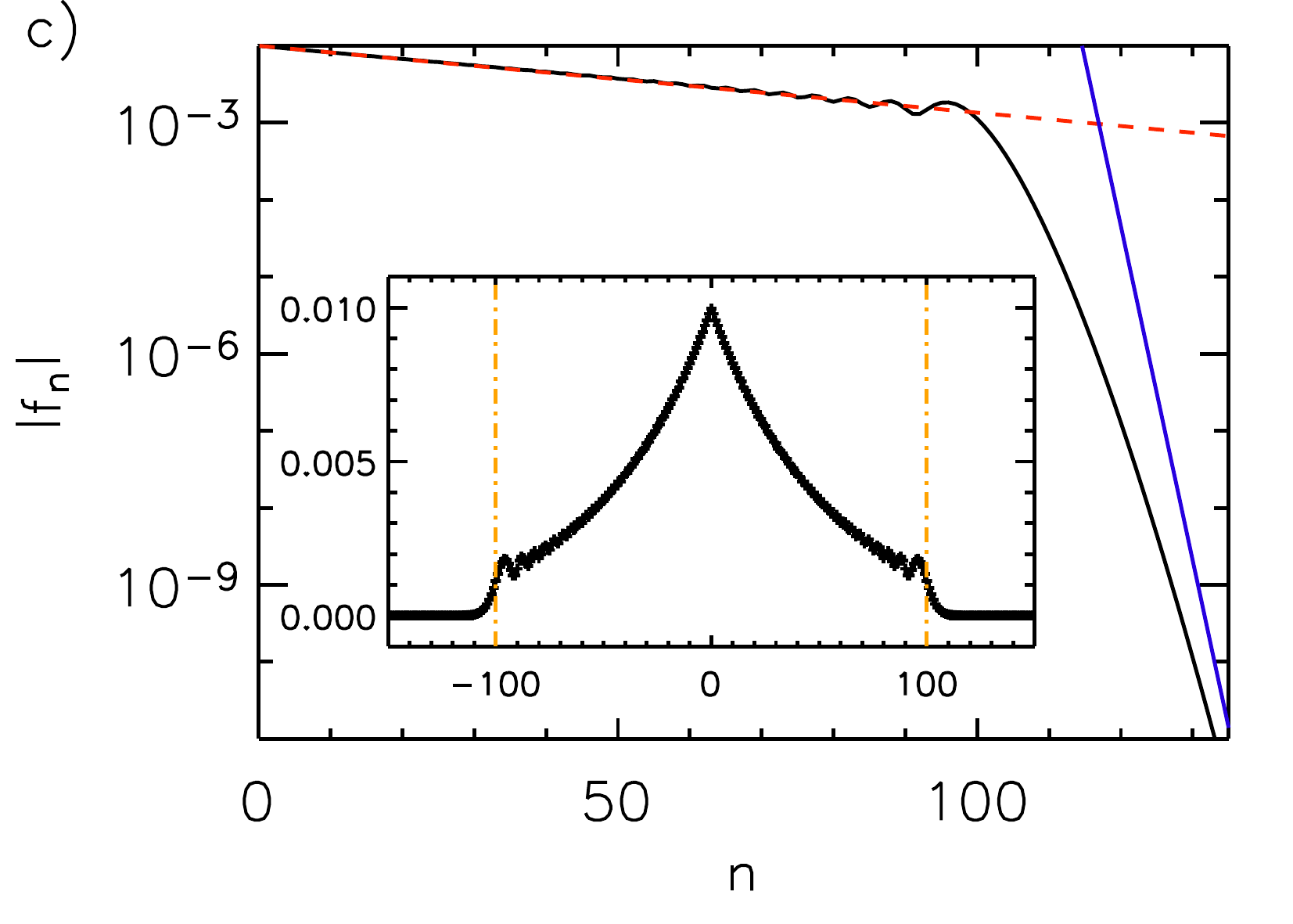}\qquad
\includegraphics[width=.45\textwidth]{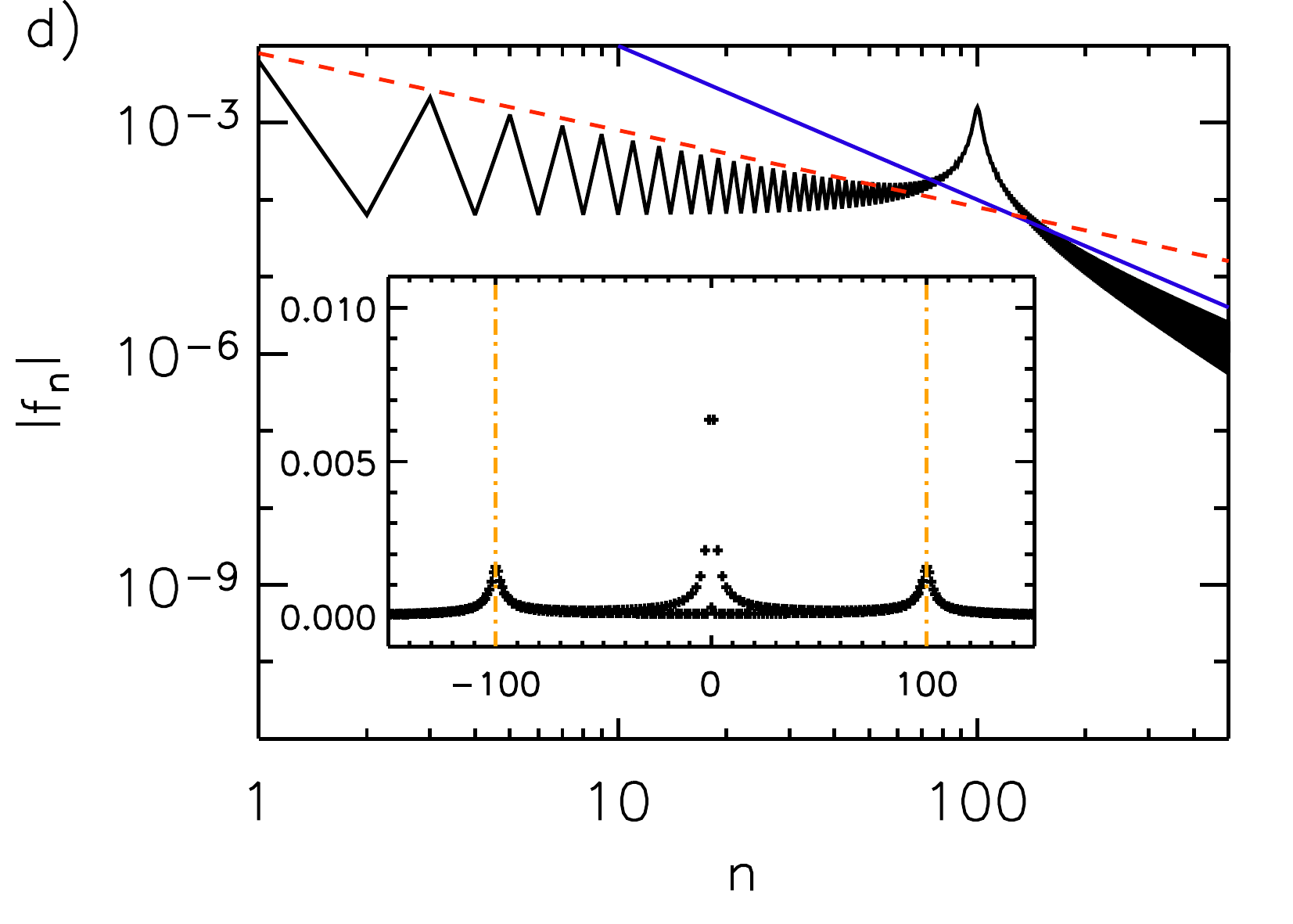}}
}
\caption{\label{fig:sbsexact}
Coefficients $|a_n|$ without external excitation with $Z/\Omega=100$ for sinusoidal (a) and rectangular (b) driving (see \eref{anSinuasoidal} and \eref{anRectangular}). The inset in (a) shows a corresponding semi-logarithmic plot, where the dashed line $\propto e^{-n}$. The $(-1)^n$ in the upper Eq.\,\eref{anRectangular} results in an odd-even staggering of the amplitudes $a_n$, for that reason the inset in (b) plots only the odd $n$ in a double-logarithmic plot, where the dashed line $\propto n^{-2}$. The resulting sideband structures $f_n$ (Eq.\,\eref{eq:f_n}) are plotted for sinusoidal (c) and rectangular (d) driving. The carrier frequency was chosen at the centre of the modulation band. The insets show the corresponding data in a linear plot. In (c) the blue solid line corresponds to $\propto e^{-n}$ and the red dashed line to $\propto e^{-2n\Omega/Z}$. In (d) the red dashed line corresponds to $\propto n^{-1}$ and the blue solid line to $\propto n^{-2}$.
}
\end{figure}

In figures \ref{fig:sbsexact}(a) and (b) the moduli of coefficients $a_n$ are shown for sinusoidal and rectangular driving for $Z/\Omega=100$. A rise of amplitudes near points $n=\pm Z/\Omega$ is found in both cases. \Fref{fig:sbsexact}(a) shows an exponential decay of amplitudes outside of the region $-Z/\Omega<n<Z/\Omega$ (see inset), while in \fref{fig:sbsexact}(b) the decay is algebraic (see inset). Using \eref{eq:f_n} the corresponding sideband structures have been calculated. They are shown in \fref{fig:sbsexact}(c) and (d). In case of the sinusoidal driving an exponential decay is observed, whereas in case of the rectangular driving an algebraic decay was found. As usual, exponential and algebraic decays of sidebands correspond to an analytic and step-wise driving respectively. One can notice a peculiar behaviour of the sideband structure close to the turning points in \fref{fig:sbsexact}\,(c). An intuitive argument explaining this behaviour was given in the introduction.

\subsection{The features of sideband structures}
\label{ssec:FeatSideband}

\begin{figure}
\begin{center}
\includegraphics[width=10cm]{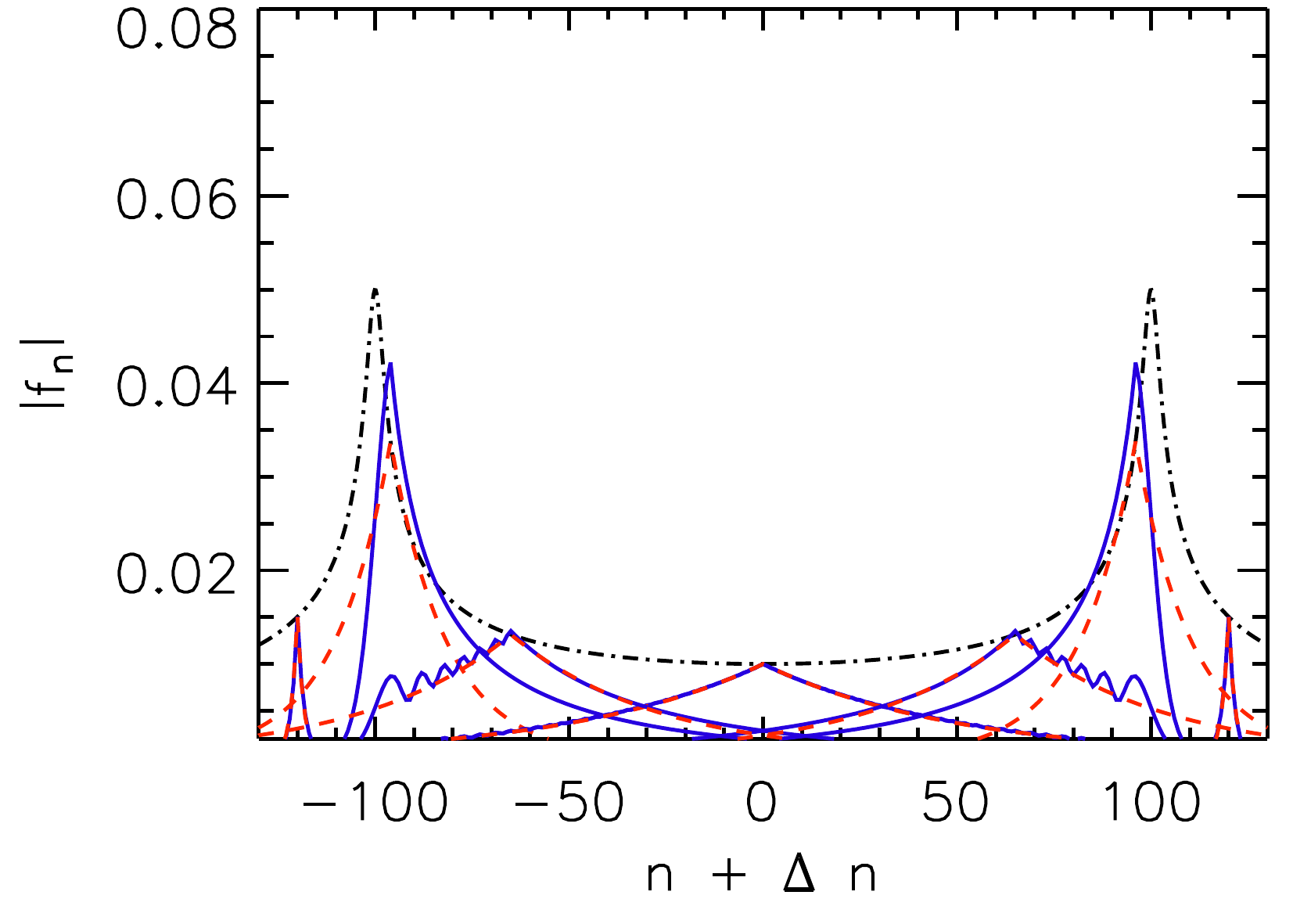}\\
\end{center}
\caption{\label{fig:sbssin}
Envelops of sideband structures $f_n$ for the sinusoidal driving with $Z/\Omega=100$, $\gamma/\Omega=2$. The blue lines correspond to \eref{eq:f_n}, the red dashed lines represent the approximation \eref{eq:f_n-2}. The dashed-dotted curve is an envelope of the main peaks \eref{eq:f_n-6}.}
\end{figure}

In the last subsection we presented an analytical solution of equation \eref{eq::12}. However this solution is not convenient for a further analysis due to the infinite sum in \eref{eq:f_n}. For this reason we will simplify the solution.

Let us first neglect the time derivative in \eref{eq::12}. Then the solution reads
\begin{equation}
f(t)=\frac{\omega_0}{\omega_c-\omega_0+d_1(t)+i\gamma}
\label{eq:31}
\end{equation}
and
\begin{equation}
f_n=\frac{\Omega}{2\pi}\int_{-T/2}^{T/2}\frac{\omega_0 e^{in\Omega t}dt}{\omega_c-\omega_0+d_1(t)+i\gamma}.
\label{eq:f_n-2}
\end{equation}

Approximation \eref{eq:31} fails when the derivative $f'(t)$ becomes large and can not therefore be neglected in (\ref{eq::12}). For an analytic signal this happens only in the case of a weak absorption if the carrier frequency lies within a modulation band. Then close to the moment $t^*$, such that $d_1(t^*)=\omega_0-\omega_c$, expression (\ref{eq:31}) is no longer valid. A precise behaviour of the function $f(t)$ close to $t^*$ is the origin of the peculiar behaviour of sidebands close to the turning points.

\Fref{fig:sbssin} shows envelopes of sideband structures for different carrier frequencies for the case of sinusoidal driving \eref{eq:sinus} and weak absorption. Sideband structures generated by \eref{eq:f_n} are compared with the approximative results \eref{eq:f_n-2}. One can see that \eref{eq:f_n-2} generates sideband structures with a smooth profile and a width dependent on the location. Within the modulation band sideband structures are broader while outside this range they are narrower. Amplitudes of harmonics beyond the classically allowed range decay much faster than those inside.

Far from the two turning points \eref{eq:f_n-2} works very well. In the vicinity of a turning point \eref{eq:f_n} produces oscillations on top of smooth profiles given by \eref{eq:f_n-2}. These oscillations appear due to the localised singularities of $f(t)$ which are not reproduced by \eref{eq:31}.

The amplitude of the main peak $f_0$ for the sinusoidal driving in the framework of \eref{eq:f_n-2} can be calculated analytically. We find
\begin{equation}
f_0=\frac{\omega_0}{\sqrt{(\omega_c-\omega_0+i\gamma)^2-Z^2}}.
\label{eq:f_n-6}
\end{equation}
In \fref{fig:sbssin} the envelope $f_0(\omega_c)$ is shown by the dashed line, which approximates also the maxima of \eref{eq:f_n} well. It only underestimates the maxima a bit close to the turning points.

\subsection{Semiclassical analysis of the spectrum}
\label{ssec:semiclassics}

The direct numerical calculation of harmonics $a_n$ for an arbitrary driving function $d_1(t)$ requires a computation of integrals from rapidly oscillating function in \eref{eq:a_n}. This leads to a hardly controllable accuracy. To avoid this problem we propose to use in practice an analytical approximation for $a_n$. The first step in the approximation is to use the stationary phase method \cite{Fed77} to compute \eref{eq:a_n}. It gives
\begin{eqnarray}
\fl
a_n=\sum_i \frac{\Omega}{\sqrt{2\pi d_1'\bigl(t_n^{(i)}\bigr)}}
\exp\left(in\Omega t_n^{(i)}+i\int_0^{t_n^{(i)}} d_1(t)dt+\frac{i\pi}{4}\textrm{sgn}\bigl(d_1'(t_n^{(i)})\bigr)\right),
\label{eq:a_n-2}
\end{eqnarray}
where stationary points $t=t_n^{(i)}$ are solutions of the equation
\begin{equation}
n\Omega+d_1\bigr(t_n^{(i)}\bigl)=0
\label{eq:stat_points}
\end{equation}
within the interval $-T/2<t_n^{(i)}<T/2$ and the summation in \eref{eq:a_n-2} should be performed over all solutions. If \eref{eq:stat_points} does not have solutions, the corresponding harmonics $a_n$ are exponentially small. In the frame of the considered approximation we can put $a_n=0$.

If $t^{(i)}_n$ lies in the vicinity of a stationary point $t^*$, where $d_1'(t^*)=0$, the denominator \eref{eq:a_n-2} becomes small and the approximation \eref{eq:a_n-2} fails. To overcome this problem we expand the exponent in \eref{eq:a_n} in Taylor series
\begin{eqnarray}
\fl
in\Omega t &+& i\int_0^t dt'd_1(t')\nonumber\\
\fl
&\simeq & in\Omega t^*+i\int_0^{t^*} dt'd_1(t')
+i(t-t^*)\bigl(n\Omega+d_1(t^*)\bigr)+\frac{i}{6}(t-t^*)^3 d_1''(t^*)
\label{eq:exp_exp}
\end{eqnarray}
and perform the integration replacing limits of integration $\pm T/2$ by $\pm\infty$. This gives
\begin{equation}
\fl
a_n=\frac{2^{1/3}\Omega}{(d_1''(t^*))^{1/3}}
\exp\left(in\Omega t^*+i\int_0^{t^*} d_1(t)dt\right)\textrm{Ai}\left(\frac{2^{1/3}\bigl(n\Omega+d_1(t^*)\bigr)}{\bigl(d_1''(t^*)\bigr)^{1/3}}\right).
\label{eq:a_n-foc}
\end{equation}
The last equality is valid provided that
\begin{equation}
\frac{2^{1/3}\bigl(n\Omega+d_1(t^*)\bigr)}{\bigl(d_1''(t^*)\bigr)^{1/3}}\lesssim 1,
\end{equation}
i.e. within the frequency range much smaller than the size of the modulation band.

The situation is very similar to the quantum-mechanical treatment of the movement of a particle in a potential well. Let us take for the sake of simplicity a harmonic oscillator potential, $V(x) = D x^2/2$. Its eigenfunctions (\fref{fig:intro}(b)) are obtained as solutions of the Schr\"{o}dinger equation
\begin{equation}
\psi_n''(x)=-\frac{2m}{\hbar^2}\left(E_n-V(x)\right)\psi_n(x)
\label{eq:SGL}
\end{equation}
where $E_n=\hbar\omega(n+1/2)$ and $\omega=\sqrt{D/m}$. The WKB approximation yields for the eigenfunctions in the classically allowed region (see e.g. section 3.2.3 of \cite{stoe99})
\begin{equation}
\psi_n(x)=\textnormal{Re }\frac{1}{\sqrt{2\pi|p_n(x)|}}e^{i\left(\int\limits_{0}^x p_n(z) dz -\frac{n\pi}{2}\right)}
\label{eq:35}
\end{equation}
where $p_n(x)=\sqrt{2m(E_n-V(x))/\hbar^2}$. Outside the classically region the eigenfunctions become exponentially small. At the classical turning points corresponding to $p_n(x)=0$ the WKB approximation fails, but just as in \eref{eq:a_n-foc} again the transition regime can be covered in terms of Airy functions. Oscillations due to the Airy functions are seen close to the turning points in \fref{fig:sbssin} (solid blue lines).

This close analogy in particular of \eref{eq:a_n-2} and \eref{eq:35} was our motivation to transfer terms like ``classically allowed'',``classically forbidden'' etc. to the Floquet system.

\section{Experimental results and comparison with the model predictions}
\label{sec:ExpResults}

\begin{figure}
\mbox{\raisebox{3.5cm}[0pt][0pt]{(a)}\hspace*{-0.35cm}
\includegraphics[width=6cm]{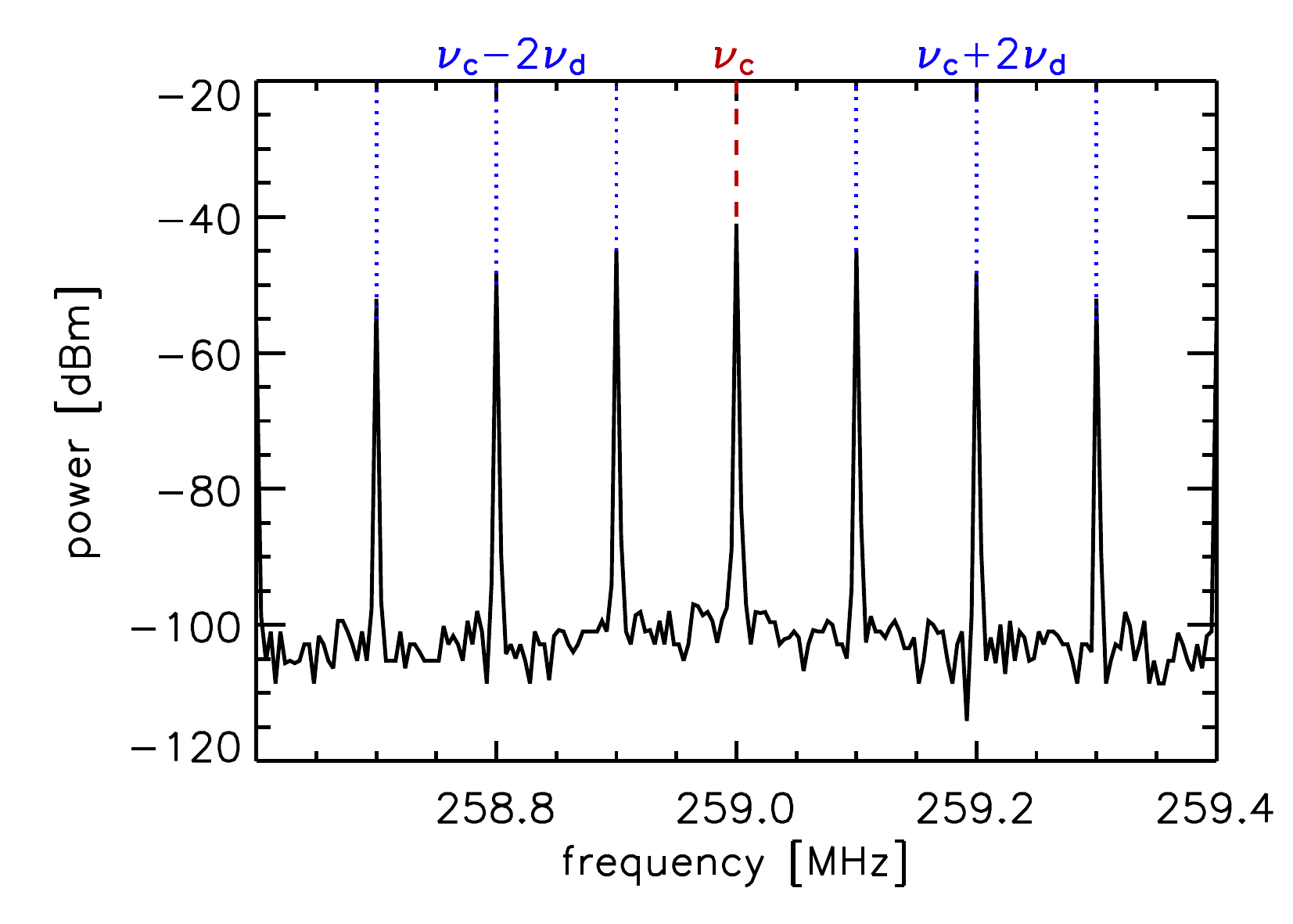}
\raisebox{3.5cm}[0pt][0pt]{(b)}\hspace*{-0.35cm}
\includegraphics[width=6cm]{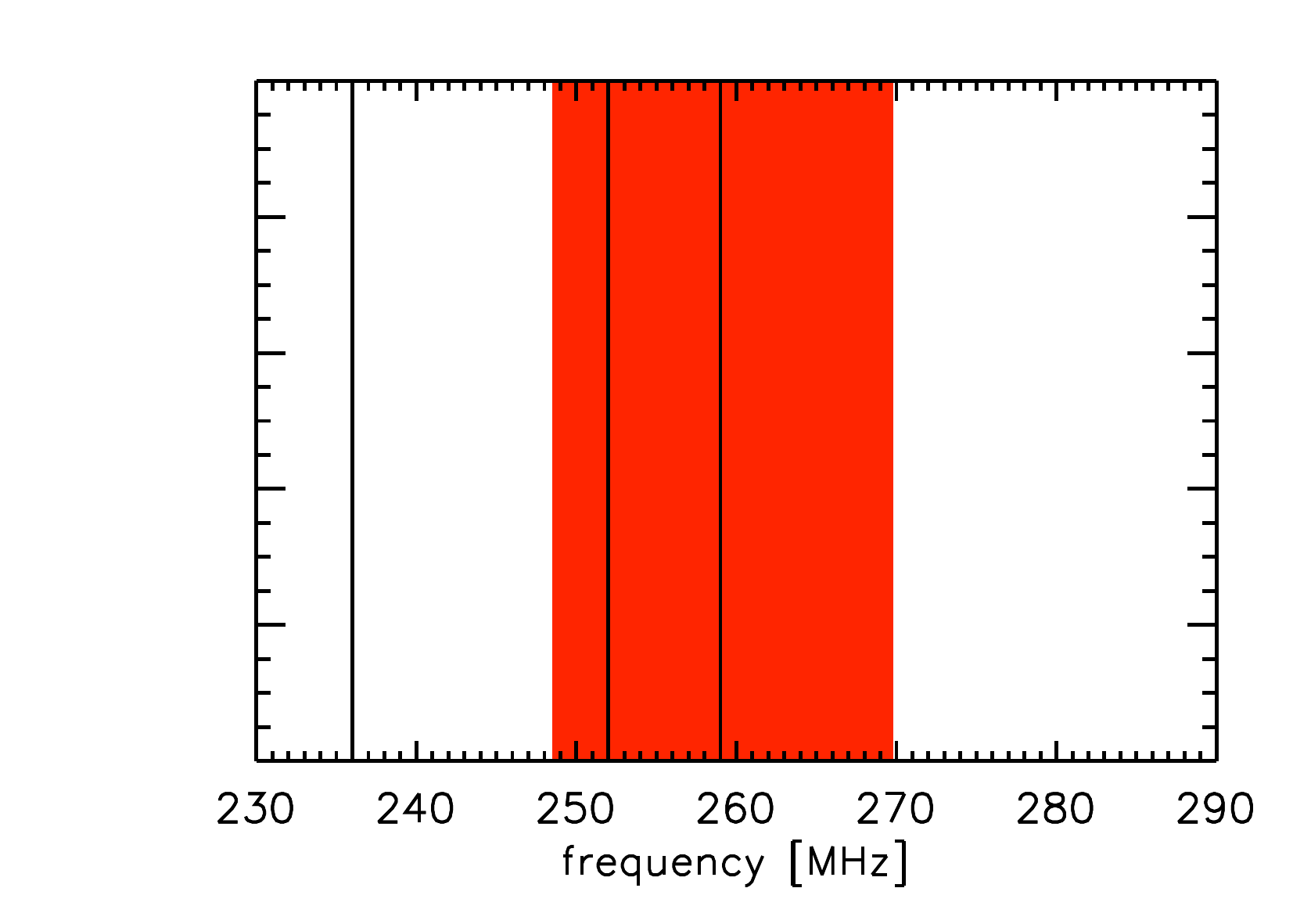}}
\caption{\label{fig:spoverview}
(a) Experimentally measured spectrum for sinusoidal driving with a driving frequency $\nu_d=\Omega/2\pi=100$\,kHz and a carrier frequency of $\nu_c=259$\,MHz is shown. The distance between the sidebands is exactly the driving frequency $\nu_d$.
(b) Sketch of the measured frequency range. Red shaded region corresponds to the range of eigenfrequencies covered by the modulation (extracted by $DC$ measurements) and vertical black lines to several values of carrier frequency $\nu_c$ used to probe the spectrum.}
\end{figure}

Now we can turn to the analysis of the experimental spectra and compare them with predictions based on Eq.\,\eref{eq::12}. \Fref{fig:spoverview}(a) shows a typical experimental sideband structure for sinusoidal driving. As expected the sidebands are equidistant with a distance corresponding to the driving frequency. The base value of $-100$\,dBm is the noise threshold of the spectrum analyser. The input power of the carrier wave is -20\,dBm for all shown experimental results. We carefully checked that a change of the input power in this regime leads only to a linear increase of all sideband structures. \Fref{fig:spoverview}(b) shows the frequency range covered by the modulation, i.e.\, the allowed region, shaded in red. For a better visualisation we include similar plots as insets in the following figures. The vertical solid lines correspond to carrier frequency values $\nu_c$ selected for the plots in \fref{fig:focal-points} and \fref{fig:RectSidebandDecay}.

In the time-dependent driving experiments we limited the AC-voltage to the range between $1.5$V\, and $2.5$V. Within this range the position of the resonance depends linearly on the voltage and the width stays approximately constant, see \fref{fig:res-pos}(b) and (c). From \fref{fig:res-pos}(b) we conclude that turning points lie at 248.5 and 269.8\,MHz.

We used several functional dependencies $U(t)$ to drive the varicap. In this paper we will discuss cases of the sinusoidal and the rectangular driving. Our theoretical model predicts a fundamental difference in the shapes of sideband structures in these two cases (see figures \ref{fig:sbsexact}(c) and (d)), as it was discussed in \sref{sec:LinRespApprox}.

 \subsection{Sinusoidal driving}
\label{ssec:ExpSinDriving}

\begin{figure}
\begin{center}
{\bf Experiments \hspace*{3cm} Numerics}\\[1ex]
\mbox{
\parbox{0.25cm}{\raisebox{4cm}[0cm][0cm]{(a)}} \hspace*{-0.5cm}
\includegraphics[width=6cm]{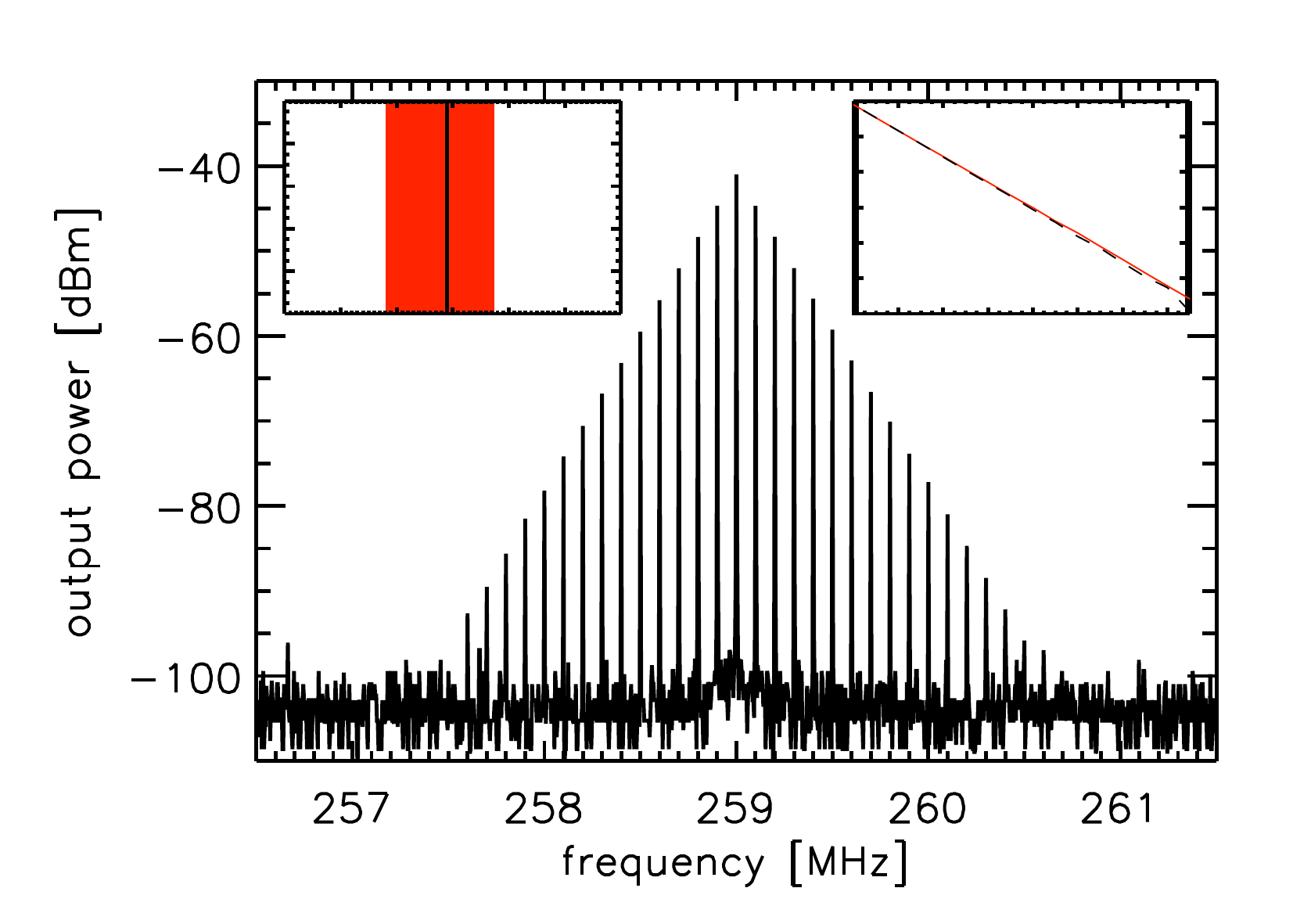}
\parbox{0.25cm}{\raisebox{4cm}[0cm][0cm]{}} \hspace*{-0.5cm}
\includegraphics[width=6cm]{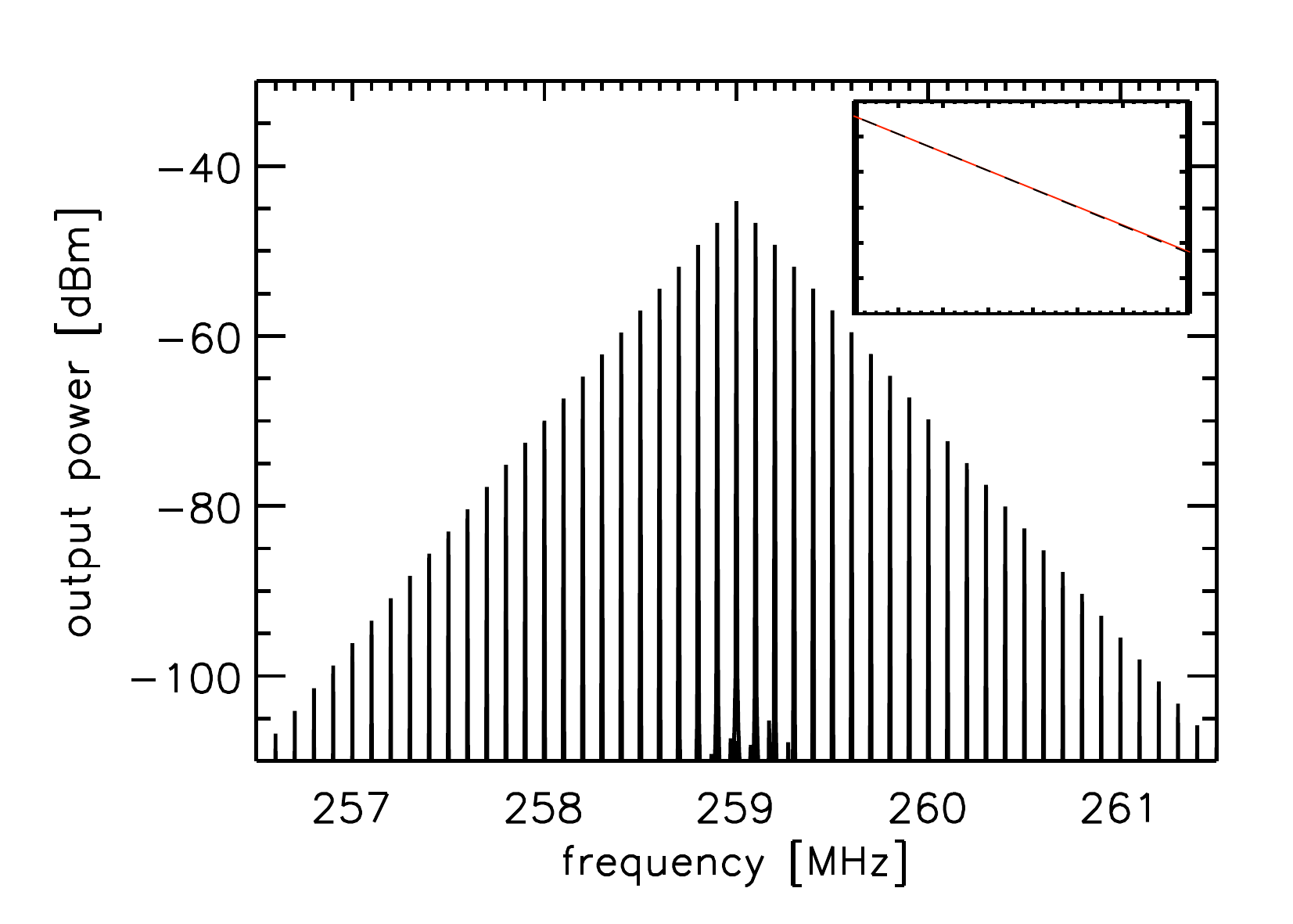}}
\mbox{
\parbox{0.25cm}{\raisebox{4cm}[0cm][0cm]{(b)}} \hspace*{-0.5cm}
\includegraphics[width=6cm]{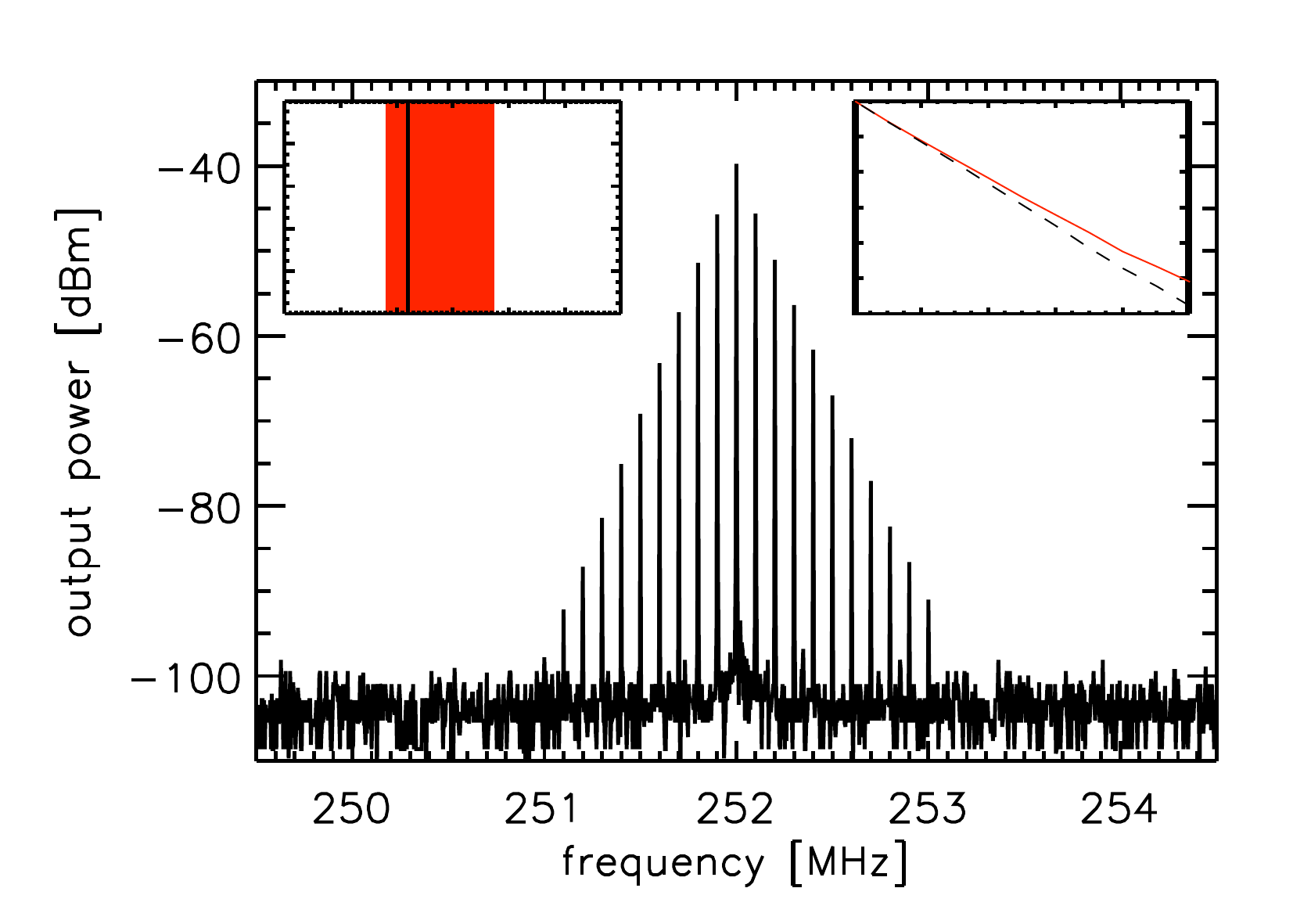}
\parbox{0.25cm}{\raisebox{4cm}[0cm][0cm]{}} \hspace*{-0.5cm}
\includegraphics[width=6cm]{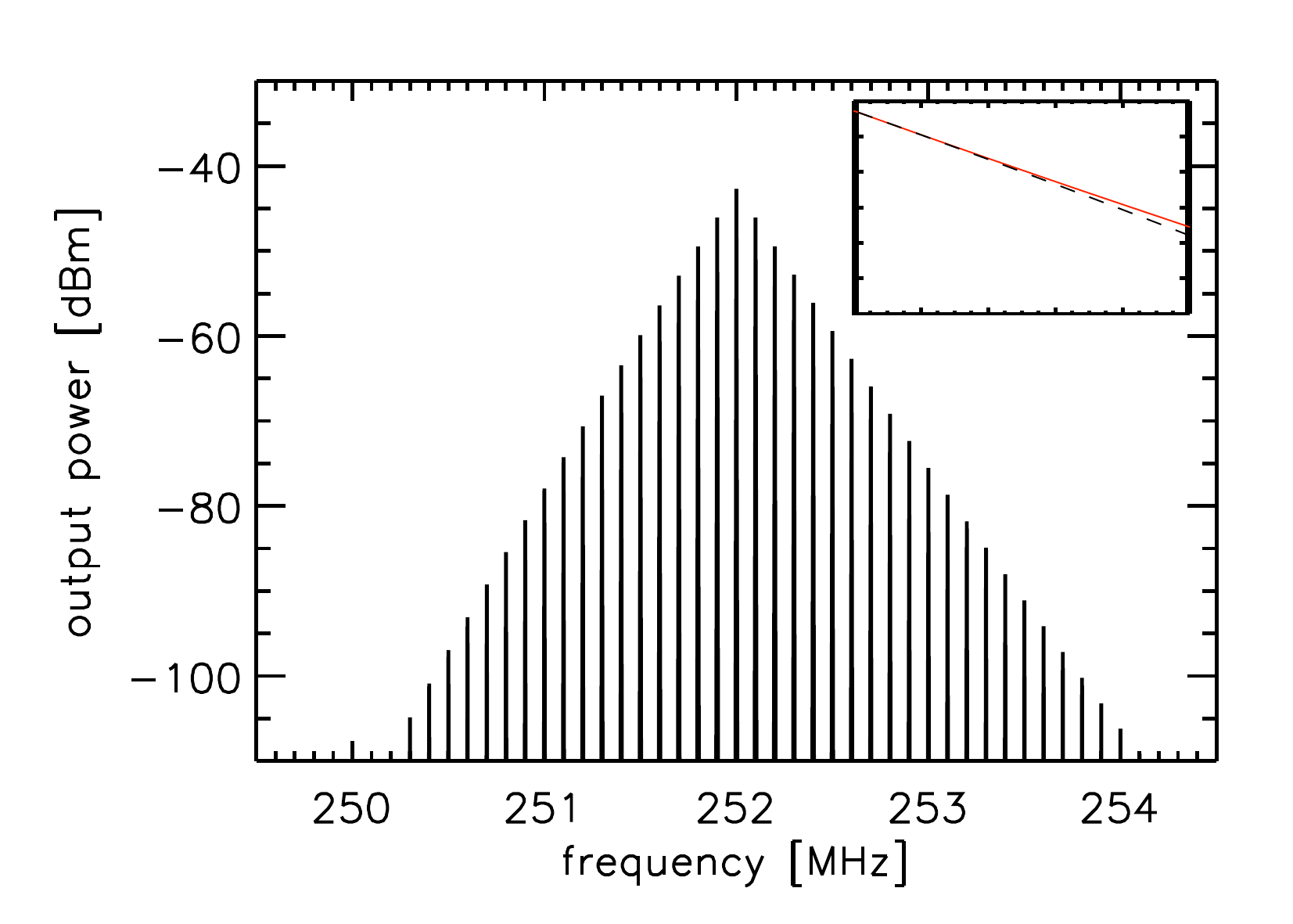}}
\mbox{
\parbox{0.25cm}{\raisebox{4cm}[0cm][0cm]{(c)}} \hspace*{-0.5cm}
\includegraphics[width=6cm]{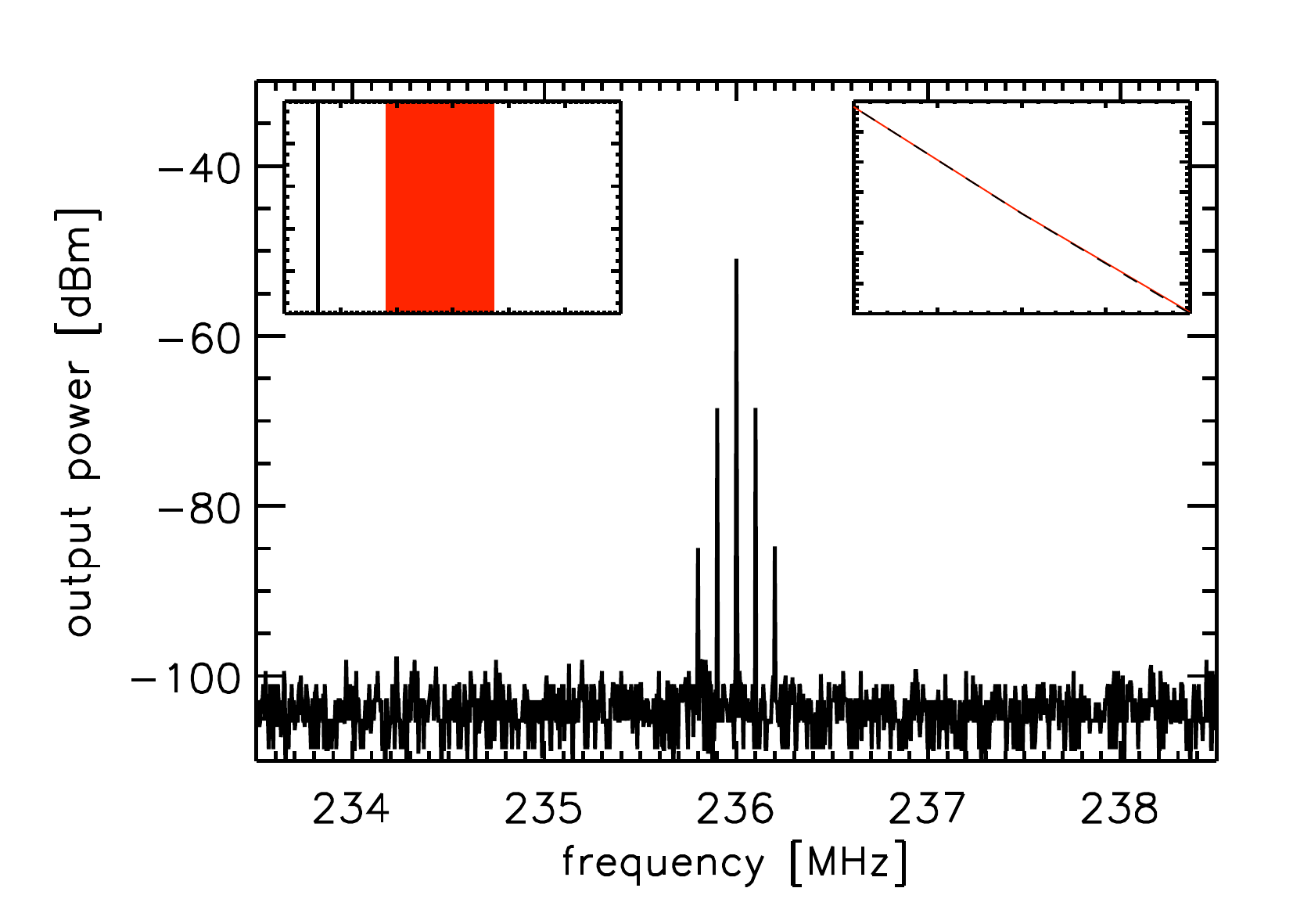}
\parbox{0.25cm}{\raisebox{4cm}[0cm][0cm]{}} \hspace*{-0.5cm}
\includegraphics[width=6cm]{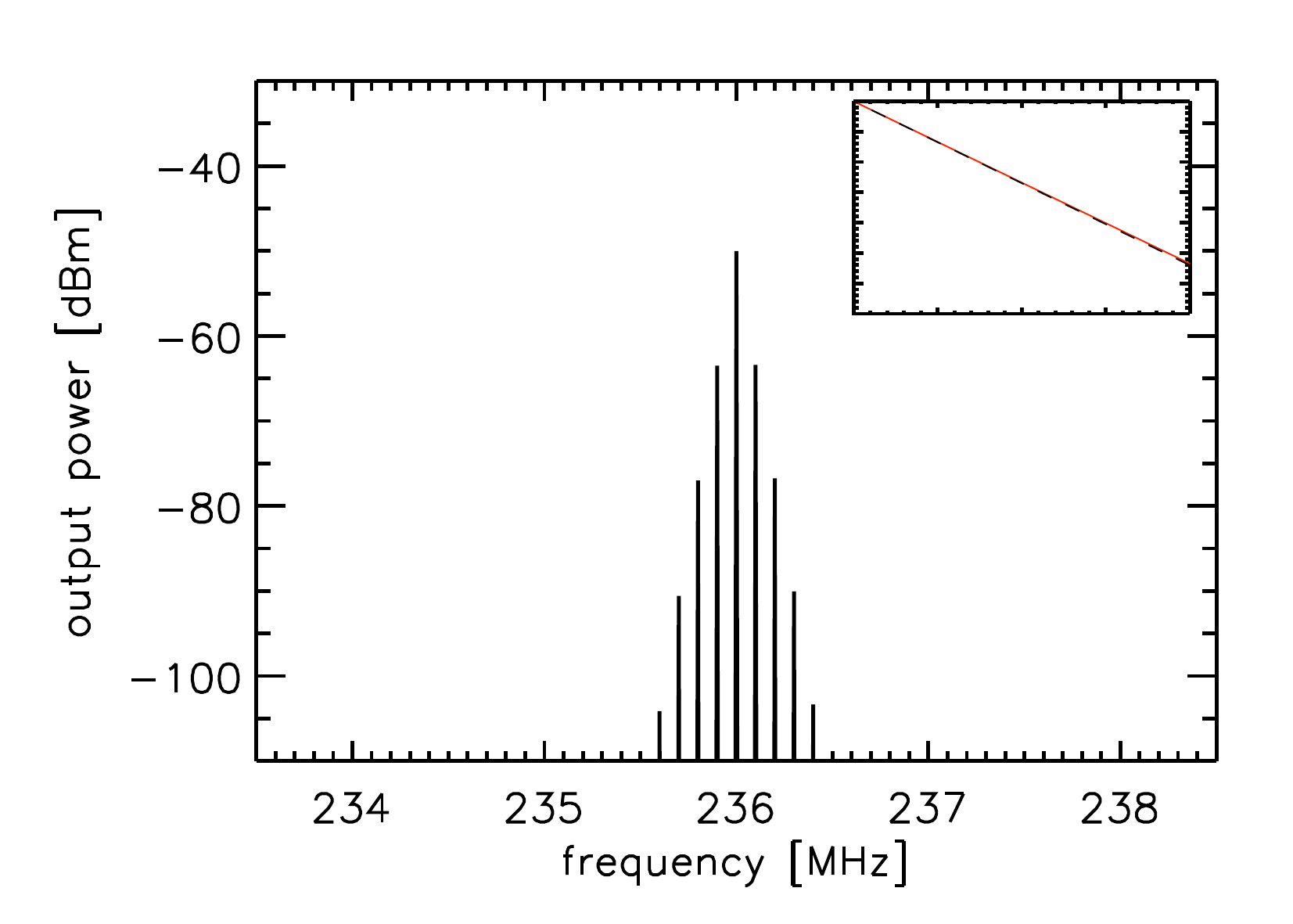}}
\end{center}
\caption{\label{fig:focal-points}
Experimental (left) and numerical (right) sideband structures observed for sinusoidal driving with $\nu_c$= 259\,MHz (a), near turning point ($\nu_c$= 252\,MHz) (b) and at $\nu_c$= 236\,MHz (c), for a driving frequency of $\nu_d=\Omega/2\pi=100$\,kHz. The left insets show the position of the driving frequency with respect to the band of the eigenfrequencies covered by the modulation. The right inset shows the envelope of the decay of the right (red solid) and left (black dashed) sidebands vs $|\nu_c-\nu_n|$ by connecting the values.
}
\end{figure}

In \fref{fig:focal-points} spectra for three different situations are shown. On the left hand side experimental results and on the right hand side numerical results are presented. For the numerics we performed a fourth-order Runge-Kutta algorithm to solve equation \eref{eq::9}. The parameters of the circuit where determined by the DC measurements (see \sref{sec:ResCircuit}). On the $y$-axis the output power is given in dBm, corresponding to a logarithmic scale. Both in the experiment and the numerics the expected exponential decay of side harmonics is clearly observed. Using the numerical solution we avoid uncertainties coming from the approximations needed to derive \eref{eq::12} from \eref{eq::9}.

For the spectra shown in \fref{fig:focal-points}(a) the carrier frequency is close to the centre of the modulation band. A zoom of this spectrum has already been shown in \fref{fig:spoverview}(a). A broad sideband structure is found in accordance with the theoretical expectation. In \fref{fig:focal-points}(b) the carrier frequency is close to one of the turning points. This gives rise to a narrower asymmetric sideband structure. The asymmetry can be seen in the inset on the right. Due to the relatively large absorption ($\sim 20\,$MHz) the observed asymmetry is small compared to the ones shown in \fref{fig:sbssin}. \Fref{fig:focal-points}(c) corresponds to the carrier frequency outside the modulation band. Here the sideband structure is very narrow. For the sake of convenience the abscissa in \fref{fig:focal-points}(a)-(c) is the same.

The comparison of experimental and numerical data shows a minor quantitative discrepancy in the decay of harmonics. This discrepancy is due to the fact that the damping $\gamma$ has been fixed by the time-independent measurement ($\sim 20\,$MHz) to have a parameter free verification, but it slightly depends on frequency. As the sideband structures are very sensitive to a change of the damping, which we will see in the next section.

\begin{figure}
\begin{center}
\mbox{
\parbox{0.25cm}{\raisebox{4cm}[0cm][0cm]{(a)}} \hspace*{-0.5cm}
\includegraphics[width=6cm]{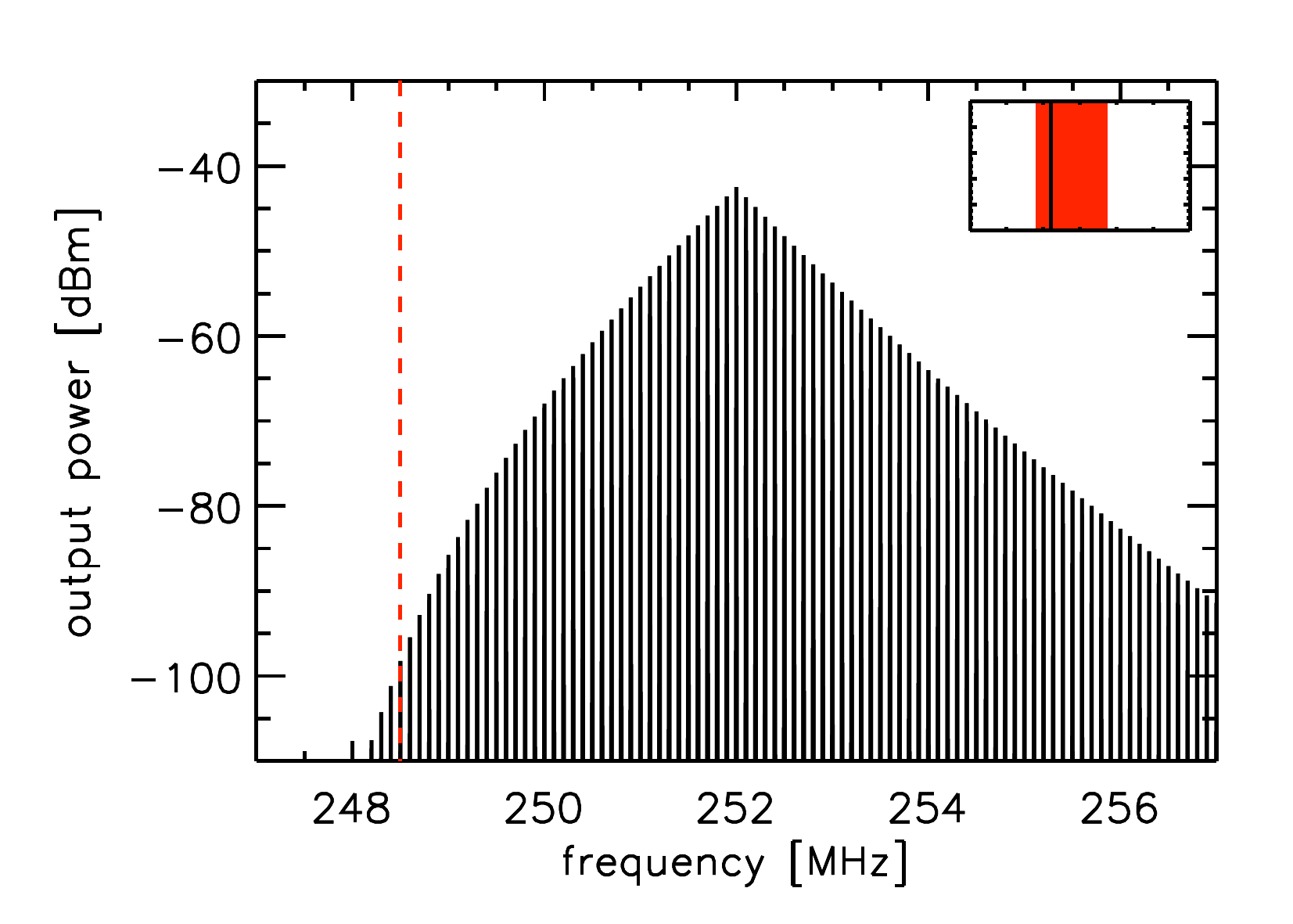}
\parbox{0.25cm}{\raisebox{4cm}[0cm][0cm]{(b)}} \hspace*{-0.5cm}
\includegraphics[width=6cm]{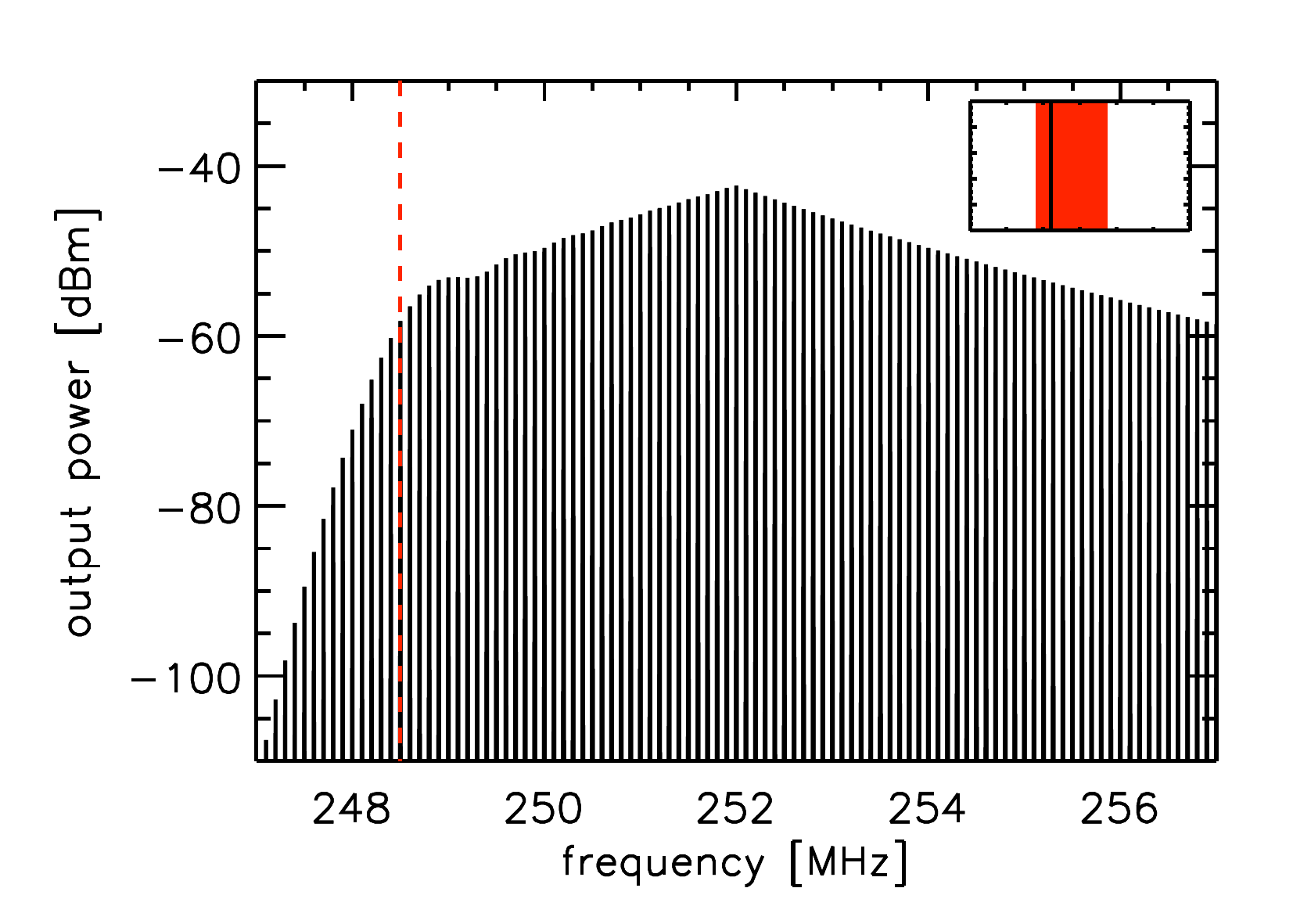}}
\mbox{
\parbox{0.25cm}{\raisebox{4cm}[0cm][0cm]{(c)}} \hspace*{-0.5cm}
\includegraphics[width=6cm]{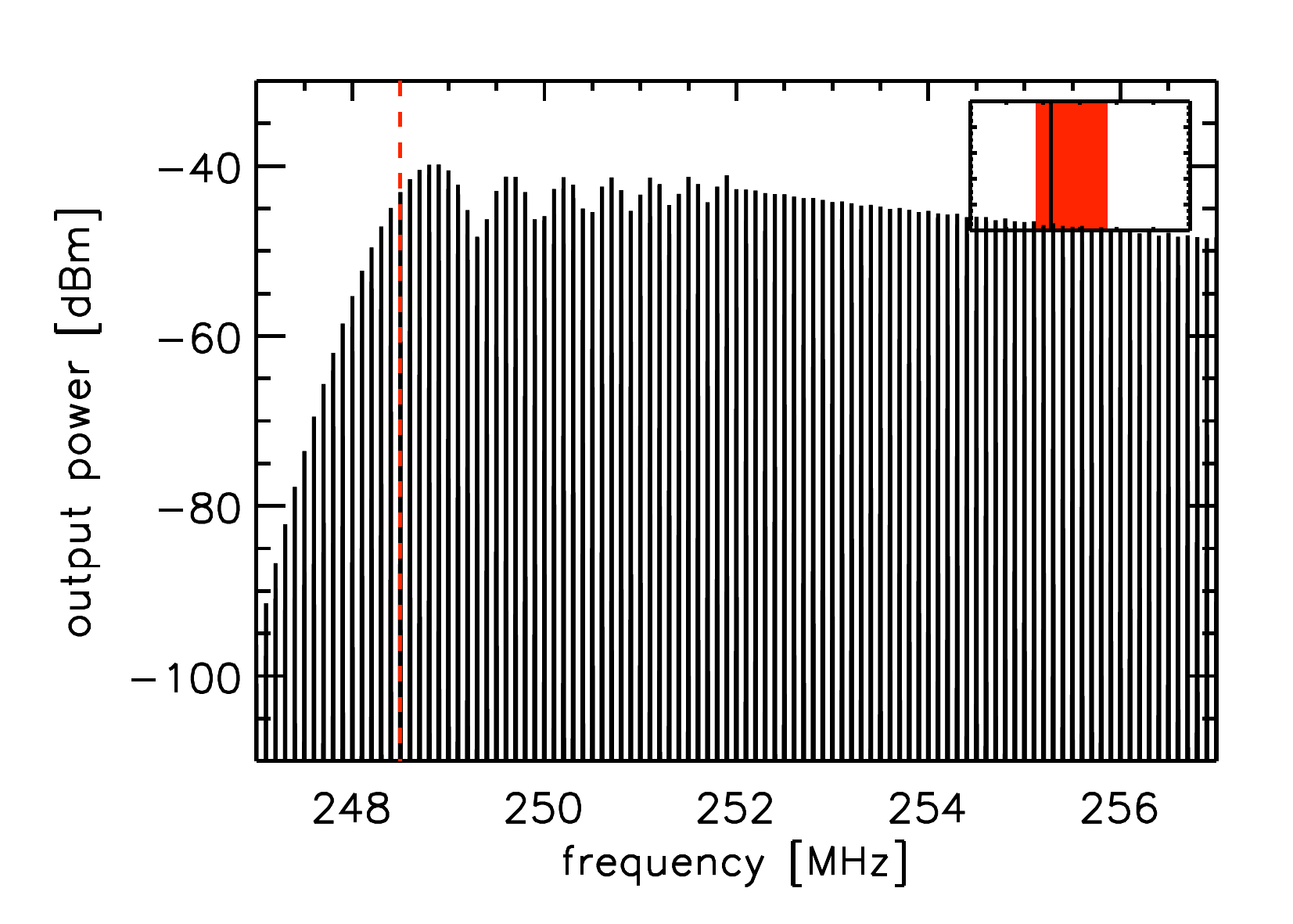}
\parbox{0.25cm}{\raisebox{4cm}[0cm][0cm]{(d)}} \hspace*{-0.5cm}
\includegraphics[width=6cm]{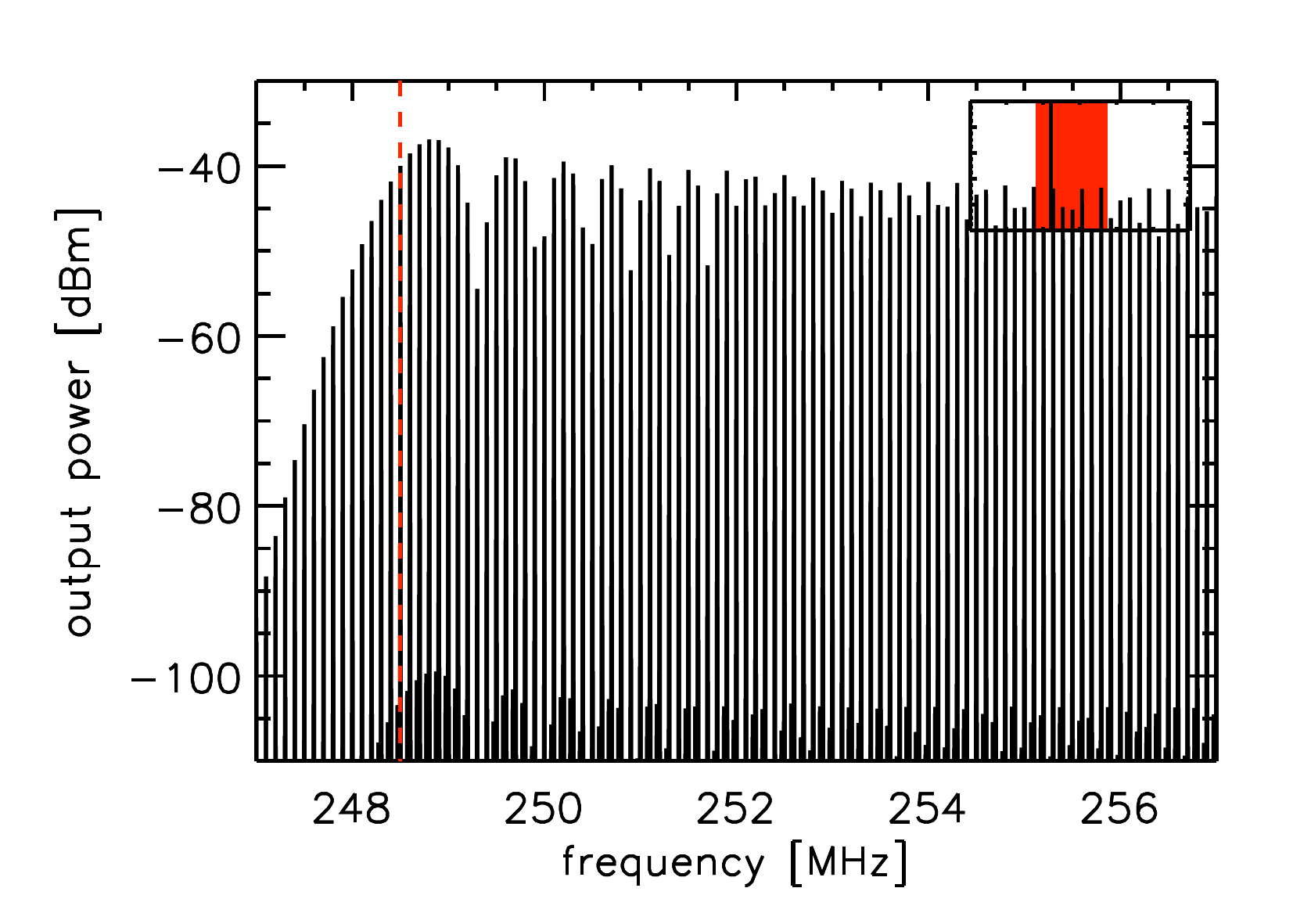}}
\end{center}
\caption{\label{fig:var-res}
Numerical sideband structures observed for sinusoidal driving with $\nu_c=252$\,MHz, $\nu_d=\Omega/2\pi=0.1$\,MHz near the turning point. From (a) to (d) the resistance is reduced so that (a) $\gamma=6.4$\,MHz, (b) $\gamma=2$\,MHz, (c) $\gamma=0.6$\,MHz, and (d) $\gamma=0.2$\,MHz.
}
\end{figure}

In \fref{fig:sbssin} we obtained an oscillating band structure close to the turning points, which is not observed in \fref{fig:focal-points}. This is an effect of the strong damping, which is illustrated by the numerics shown in \fref{fig:var-res}. The damping is reduced by a factor of three from \fref{fig:var-res}(a) to (d), respectively. The reduction of the damping transforms the slightly asymmetric triangle-like structure (\fref{fig:var-res}(a)) to a very asymmetric one with a periodic modulation of the shape close to the turning point (\fref{fig:var-res}(d)). We were unable to resolve these oscillations experimentally due to the significant minimal damping, that characterises our setup. The oscillations correspond to the Airy behaviour of \eref{eq:a_n-foc} which are also seen in \fref{fig:sbssin} close to the turning points (solid blue lines).

\begin{figure}
\begin{center}
\parbox{8cm}{\includegraphics[width=8cm]{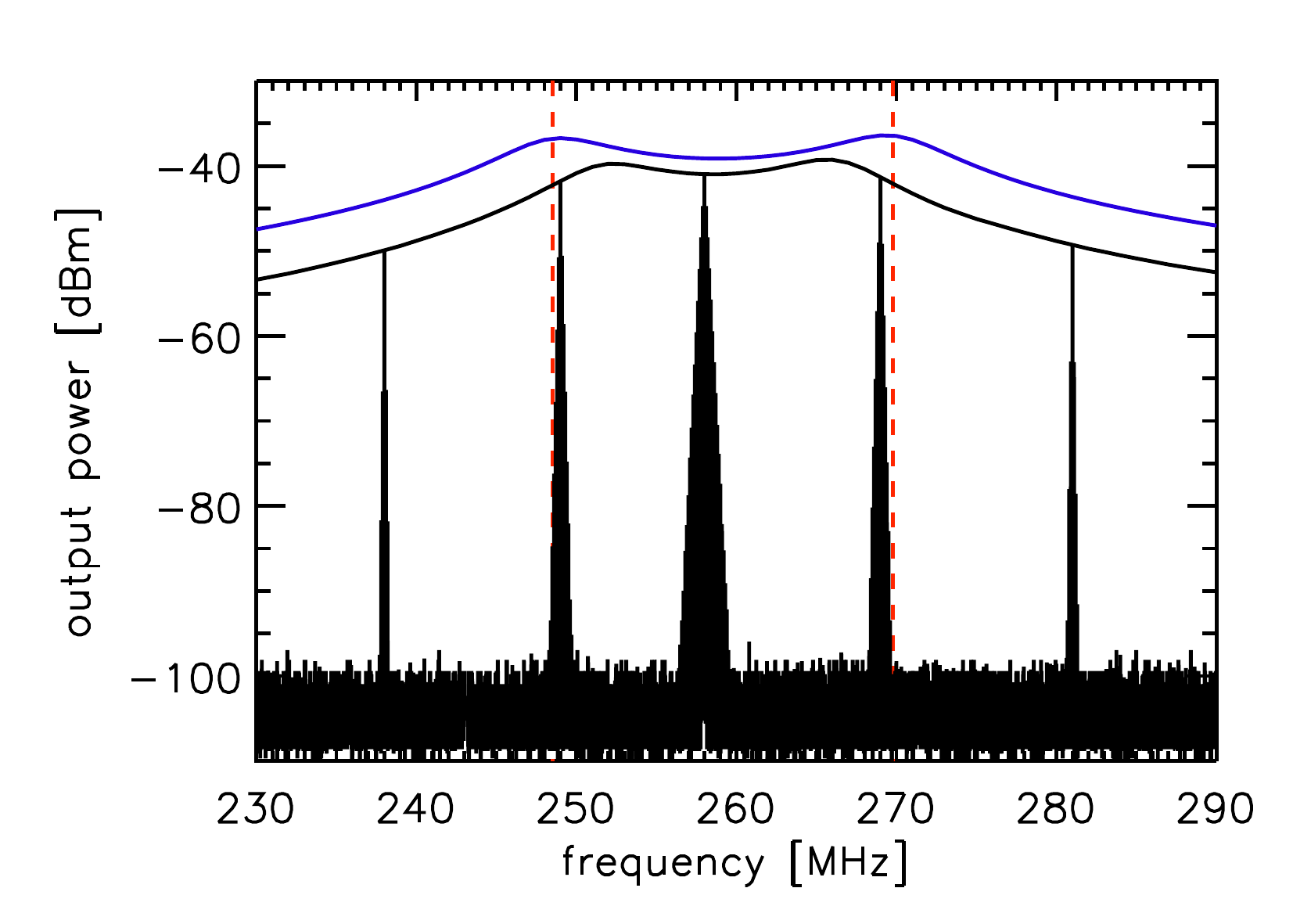}}
\end{center}
\caption{\label{fig:sin_overlay}
Superposition of sideband structures for five different carrier frequencies (sinusoidal driving as in \fref{fig:focal-points}). The black solid line corresponds to the envelop of the amplitude of the main peaks. The red dashed lines correspond to the turning points, the blue solid line to the envelop of the amplitude of the main peaks in the numerical calculation}
\end{figure}

In \fref{fig:sin_overlay} experimental sideband structures for five different carrier frequencies are superimposed. The solid black line is the envelope of the amplitude of the main peaks. The distance between main peaks was chosen to be $1$\,MHz. The blue curve is the envelope found numerically. Red dashed lines illustrate the turning points determined from the DC measurement (see \fref{fig:res-pos}\,b). The figure shows the highest amplitude of the main peak close to the turning points. This can be traced back to the divergency of the semiclassical wave function at the turning point, as already discussed in detail in \sref{sec:LinRespApprox}. Since the envelope is symmetric with respect to the centre of the modulation band the assumption that the absorption is approximately constant in case of the dynamic measurement is validated.

\subsection{Rectangular driving}
\label{ssec:ExpRectDriving}

The realization of a rectangular driving was limited by the unavoidable distortion of the signal on the slopes at high driving frequencies. Therefore experiments have been conducted at comparatively low frequency $\nu_d=\Omega/2\pi=1$\,kHz.

Due to the low frequency solving numerically \eref{eq::9} by Runge-Kutta is leading to two different time scales. One needs small time steps due to the fast carrier frequency and long time intervals to simulate slow driving frequency. Therefore numerical calculations are time consuming and inaccurate. Fortunately it is possible for the rectangular driving to solve \eref{eq::9} for the two different applied voltages individually. The solution of each harmonic oscillator needs only to be matched at the switching times, i.e.\, leading only to a boundary value problem.

\begin{figure}
\begin{center}
\mbox{
\parbox{0.25cm}{\raisebox{4cm}[0cm][0cm]{(a)}} \hspace*{-0.5cm}
\includegraphics[width=6cm]{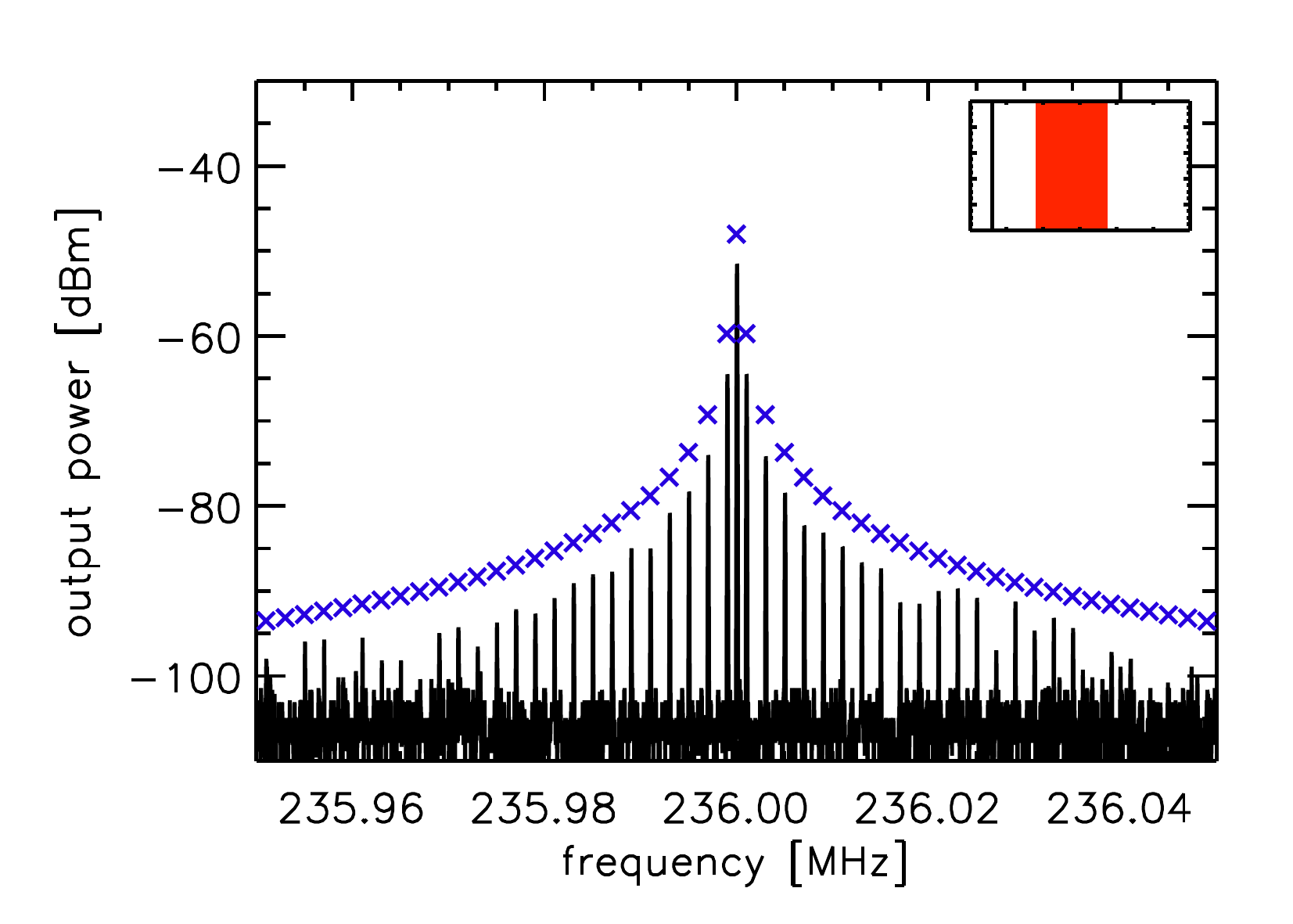}
\parbox{0.25cm}{\raisebox{4cm}[0cm][0cm]{(b)}} \hspace*{-0.5cm}
\includegraphics[width=6cm]{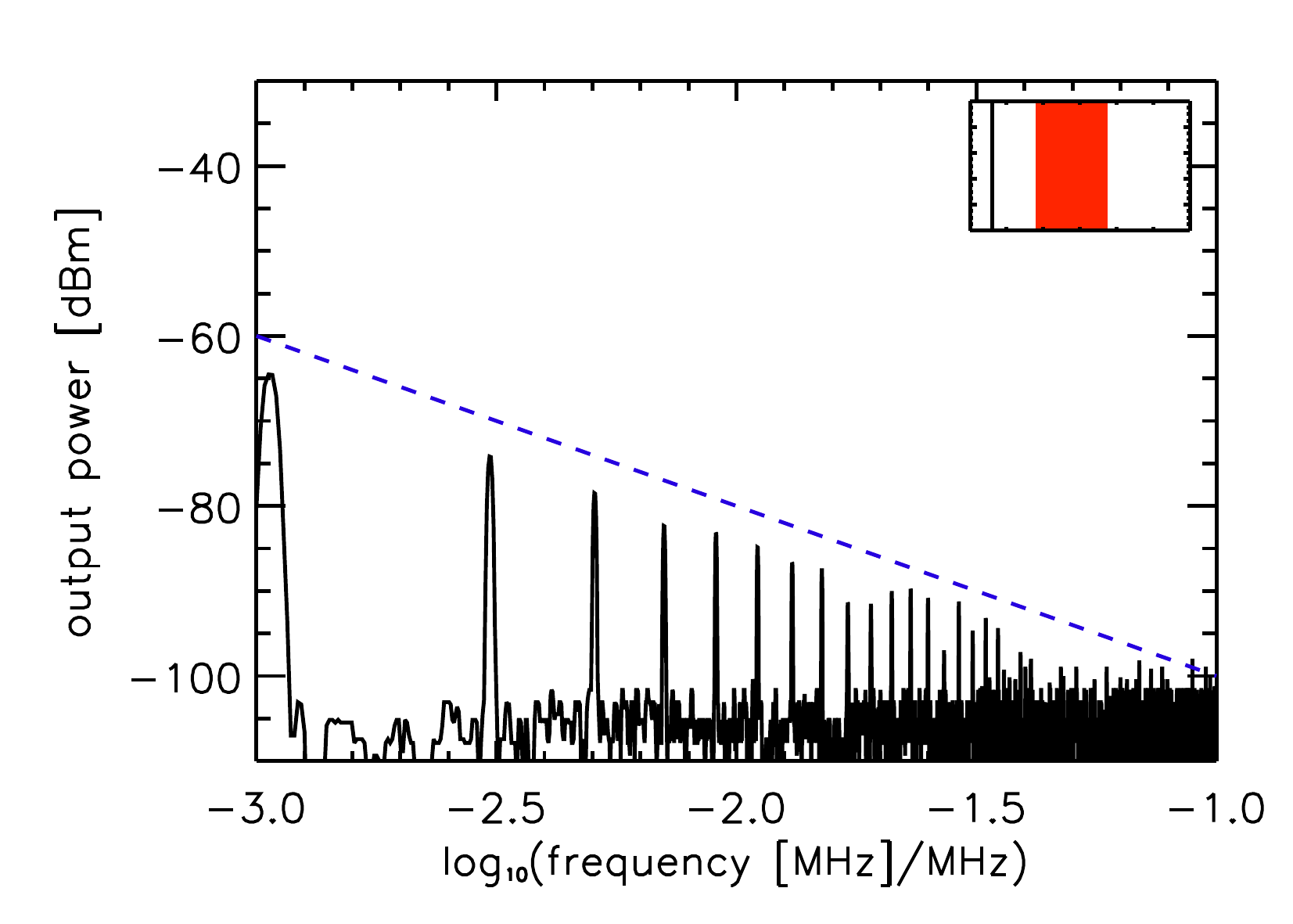}}
\mbox{
\parbox{0.25cm}{\raisebox{4cm}[0cm][0cm]{(c)}} \hspace*{-0.5cm}
\includegraphics[width=6cm]{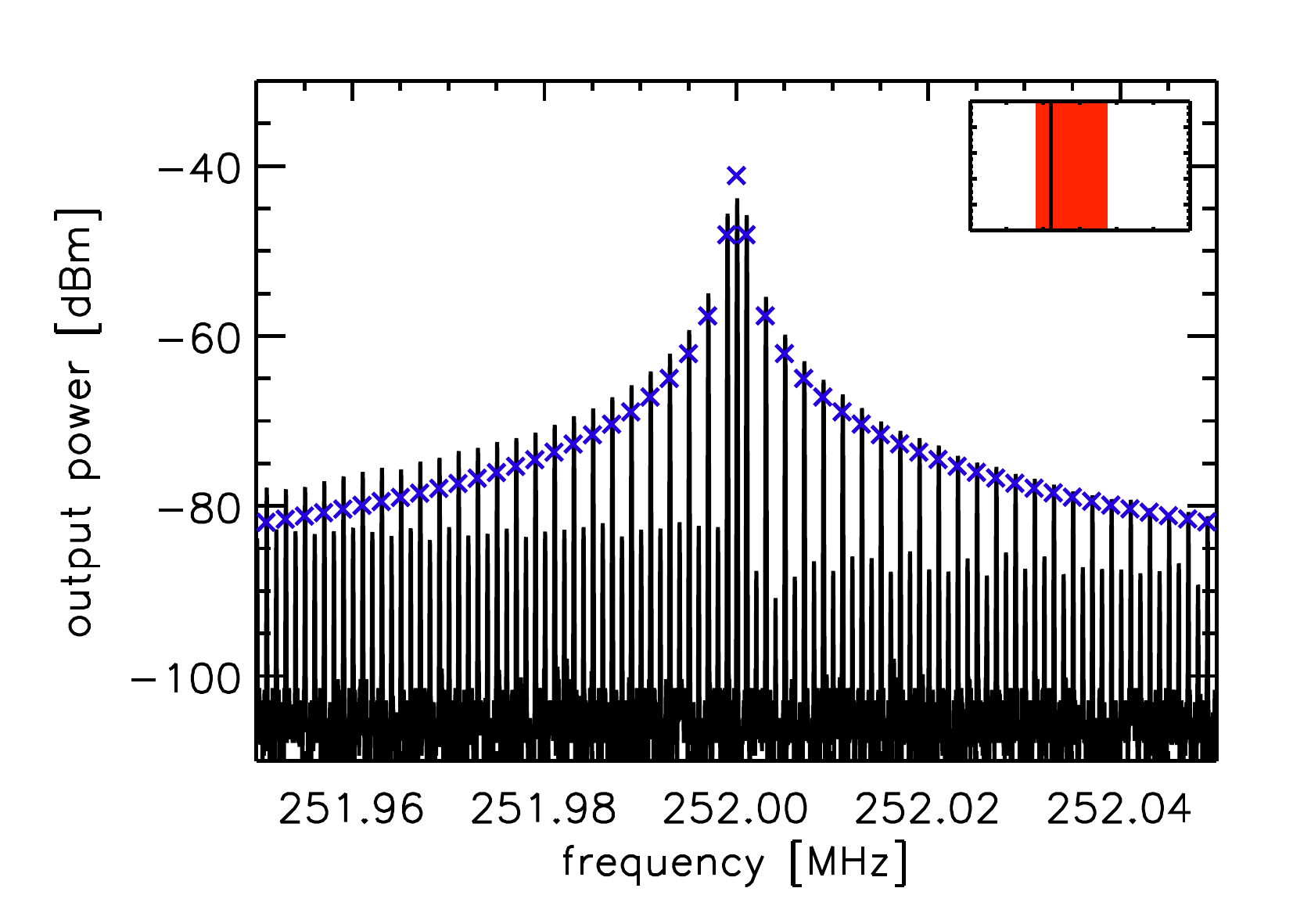}
\parbox{0.25cm}{\raisebox{4cm}[0cm][0cm]{(d)}} \hspace*{-0.5cm}
\includegraphics[width=6cm]{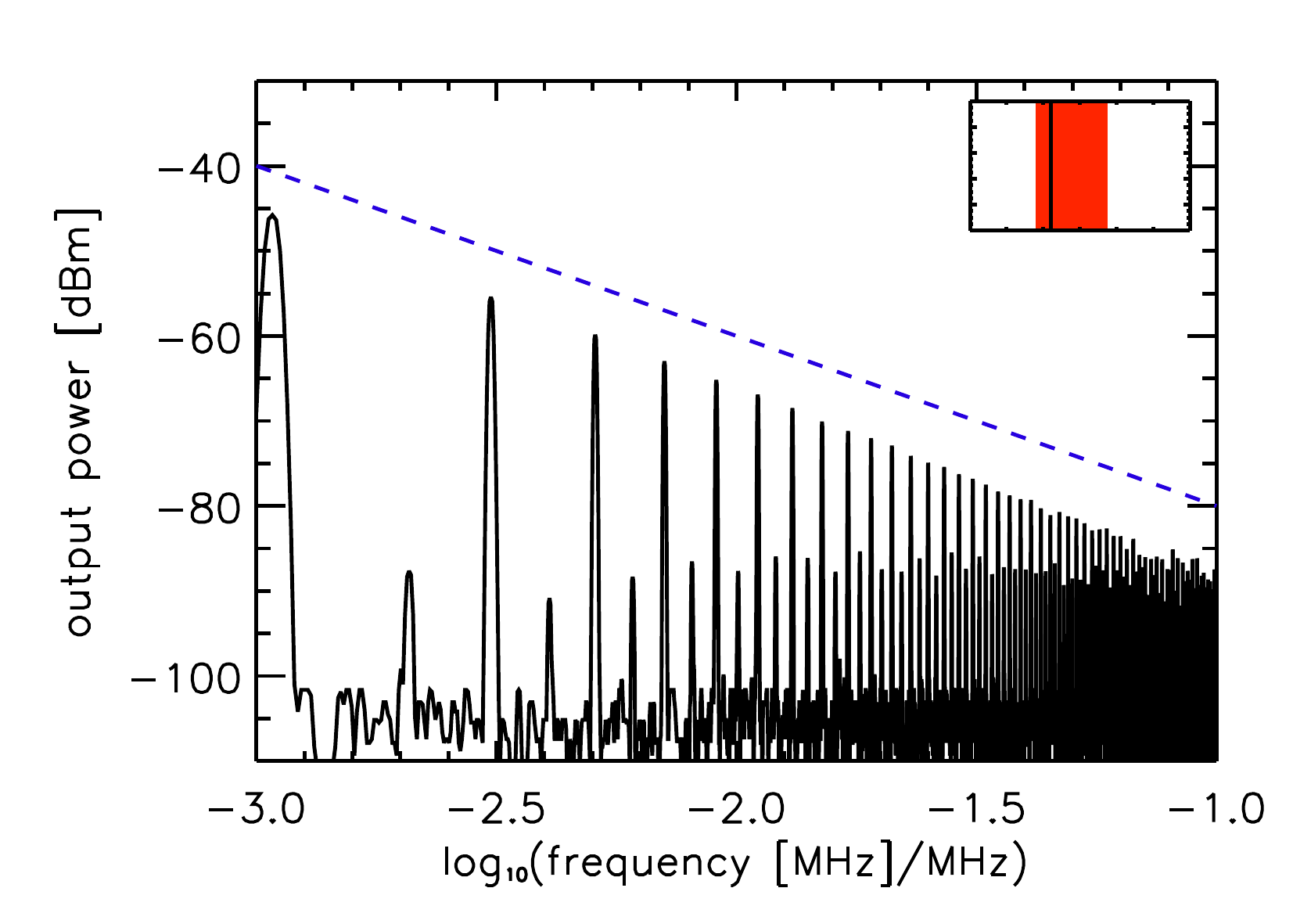}}
\mbox{
\parbox{0.25cm}{\raisebox{4cm}[0cm][0cm]{(e)}} \hspace*{-0.5cm}
\includegraphics[width=6cm]{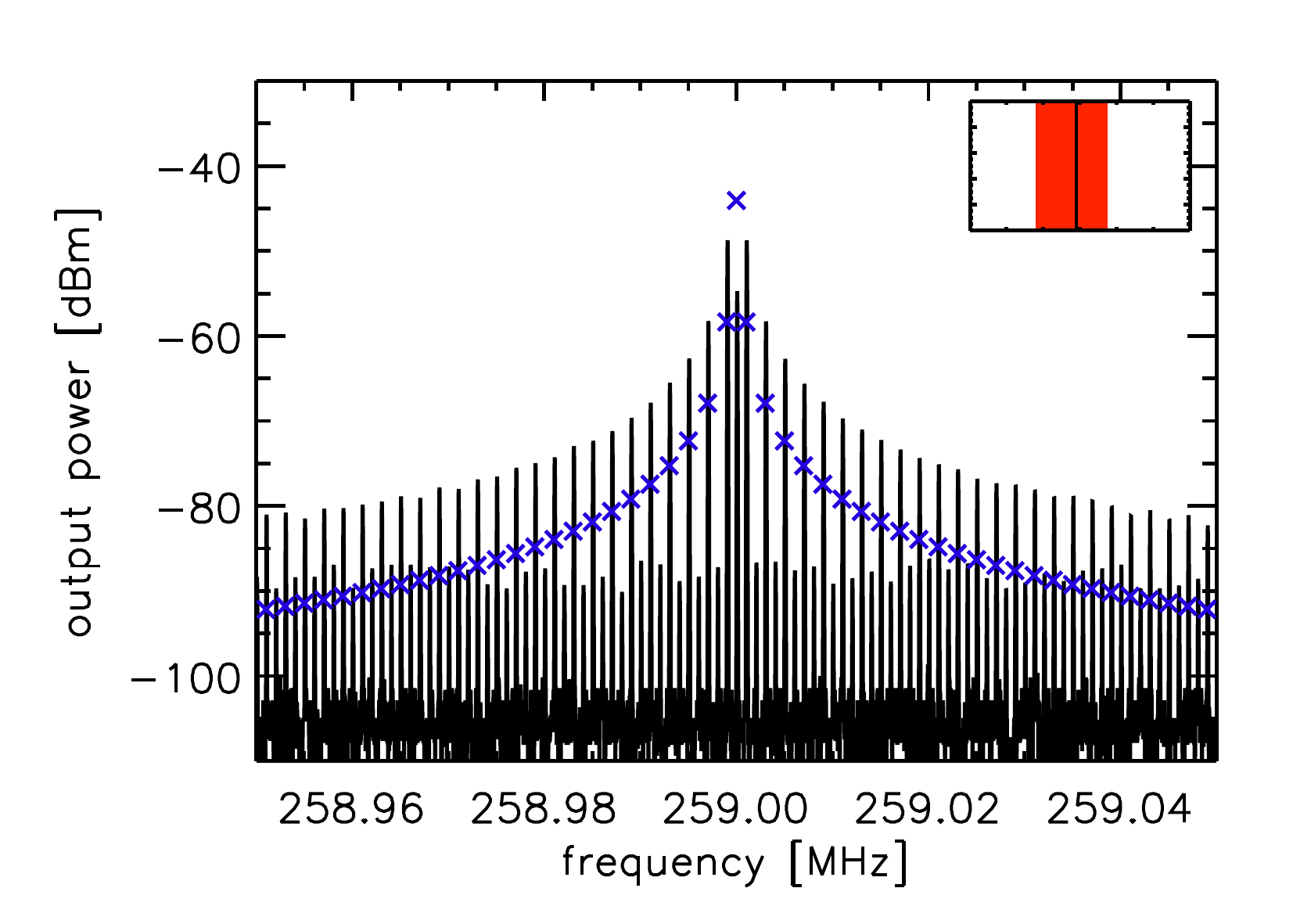}
\parbox{0.25cm}{\raisebox{4cm}[0cm][0cm]{(f)}} \hspace*{-0.5cm}
\includegraphics[width=6cm]{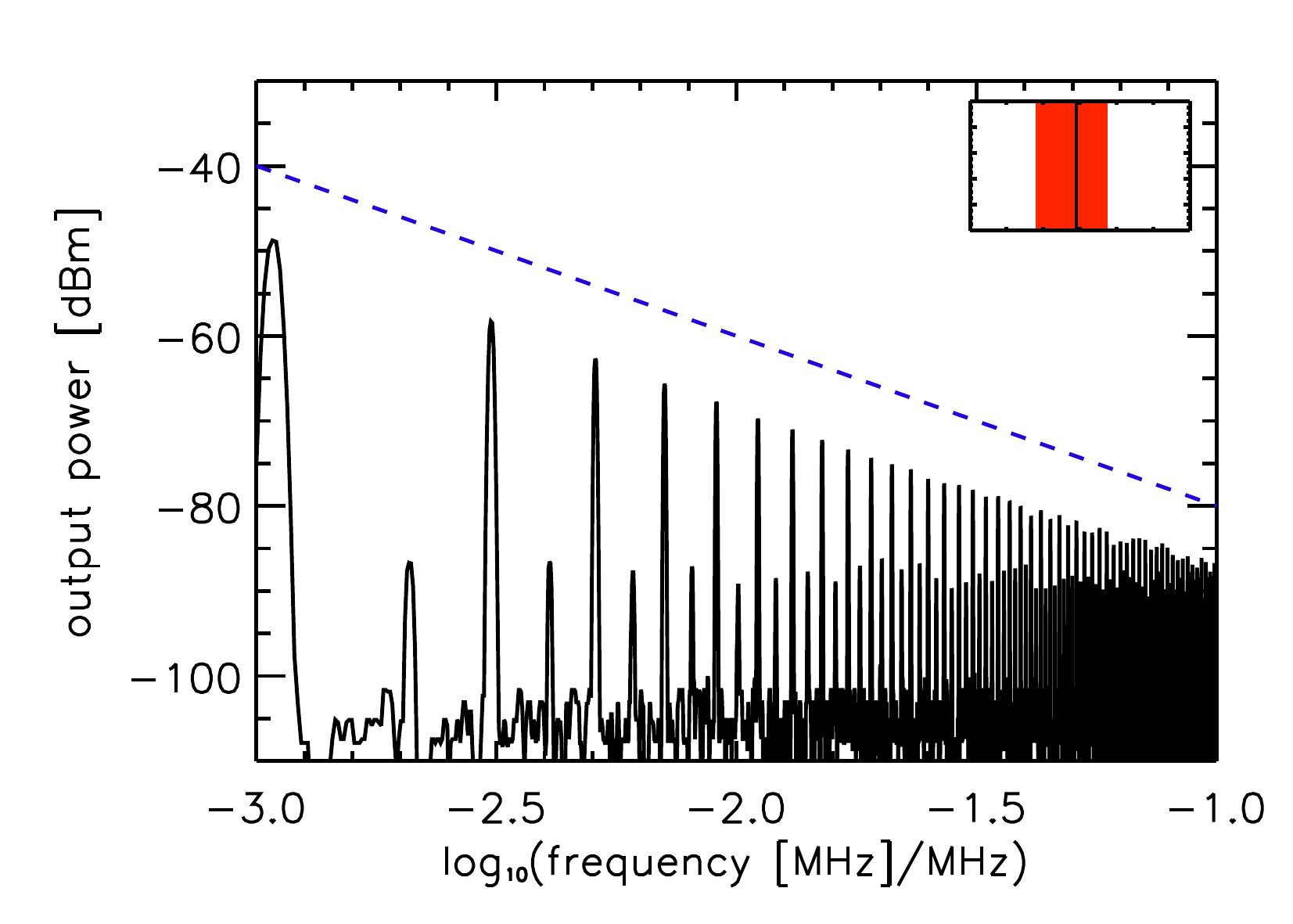}}
\end{center}
\caption{\label{fig:RectSidebandDecay}
Experimentally measured sideband structure for rectangular driving are shown in logarithmic (left) and double logarithmic (right) scale. The blue crosses corresponds to the weights of the Fourier components obtained from the analytic solution of \eref{eq::9}. On the right side the blue dashed lines corresponds to a decay with an exponent of -2\\}
\end{figure}

In figures \ref{fig:RectSidebandDecay}(a)-(f) sideband structures for three different carrier frequencies are plotted. The right spectra are plotted on a logarithmical frequency scale, allowing to verify the expected algebraic decay. In \fref{fig:RectSidebandDecay}(d) and (f) the power of the decay was found to be $-2$ in accordance with our theoretical prediction. In the left column the theoretical values of Fourier components are marked by the blue crosses. One can see a good qualitative agreement, but discrepancies in quantity. Besides the fact that the damping is extracted in the same way like in the sinusoidal driving case, the signal generator is not able to generate a precise rectangular signal. Another explanation of the observed discrepancies can lie in the validity of the simplified circuit in \fref{fig:Dose}(d) when this description is applied to the rectangular driving.

\begin{figure}
\begin{center}
\parbox{8cm}{\includegraphics[width=8cm]{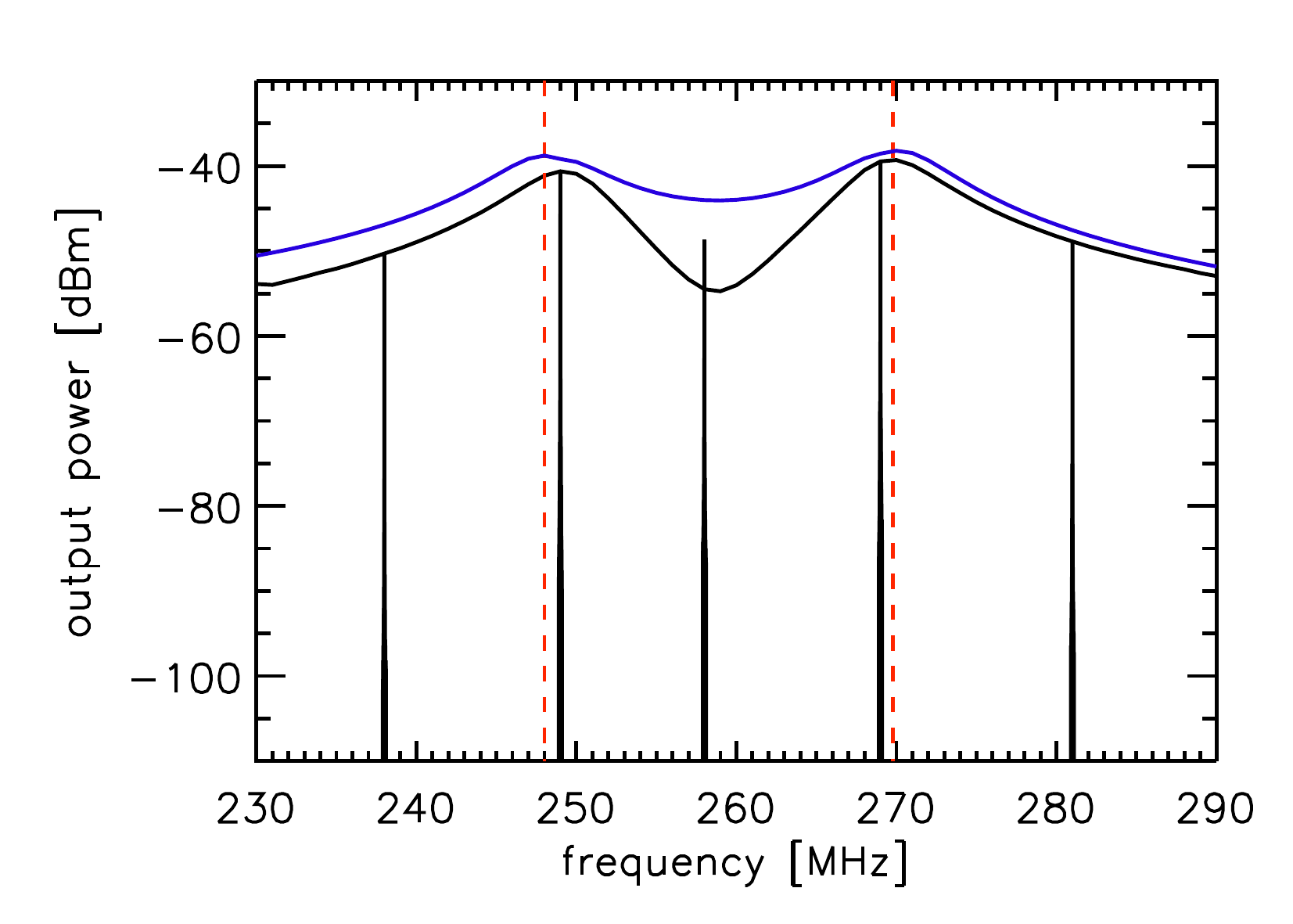}}
\end{center}
\caption{\label{fig:rect_overlay}
Overlay of sideband structures for different carrier frequencies (rectangular driving as in \fref{fig:RectSidebandDecay}), the black solid line corresponds to the envelop of the amplitude of the main peaks. The red dashed lines corresponds to the turning points, the blue solid line to the envelop of the amplitude of the main peaks for the analytical calculation}
\end{figure}

The amplitude of the main peaks (see \fref{fig:rect_overlay}) shows a behaviour like in the sinusoidal driving case, but now the minimum between the turning points is significantly deeper than in the sinusoidal case. The main peak of the sideband structure may even be smaller than the first harmonic. This was also observed in the numerics, though only rarely, and is not contained in the approximative analytic results. Experimentally even harmonics are substantially suppressed. Such an observation is in a good agreement with theory. The quantitative discrepancy may result from the non-perfect rectangular signal.

\section{Conclusions}
\label{sec:conclusion}

In this paper we have reported the realisation of the tunable microwave Floquet system. The properties of the generated sideband structures were understood within the model of a resonance circuit with a single excited resonance. Theoretically we have shown that close to the borders of the modulation band sideband structures can display large oscillations. However due to rather strong resistance of the current setup such structures have not been yet verified experimentally.

The present studies were restricted to a particular simple situation, namely one resonance circuit with just one eigenfrequency. To study questions like dynamical localisation, as mentioned in the introduction, one has to extend the studies to systems with a high density of states within the modulation band. One option is to couple several resonance circuits, like the one studied in the present work, to a microwave resonator with a sufficiently large density of states. The comparatively large resistance of the tin cup should not affect the observation of dynamical localization, as the observed algebraic decay for sidebands generated by rectangular driving is suited to show a transition to an exponential localization in the time domain. Another possible application might be a of control transport through the microwave resonator by changing the driving \cite{pro05b}.

Another promising option is to study non-linear effects. We noticed already deviations from the linear regime when increasing the power of the carrier signal. They are caused by a non-linear voltage-current characteristic of the varicap, giving rise to additional terms in \eref{eq::9}, which are similar to the non-linear terms of non-linear Schr\"{o}dinger equations. Additionally this would give the possibility to investigate the interplay between localization and non-linearity.

\section{Acknowledgements}
\label{sec:acknowledgements}

We gratefully acknowledge useful discussions with Maksim Miski-Oglu. This work was supported by the Deutsche Forschungsgemeinschaft via an individual grant and via the research group 760 ``Scattering systems with complex dynamics''.

\section*{References}

\end{document}